\documentclass[a4paper, 11pt]{article}
\pdfoutput=1
\usepackage{jheppub}
\usepackage[colorlinks = true, citecolor = blue, linkcolor = blue, urlcolor = blue]{hyperref}
\usepackage[nameinlink,noabbrev]{cleveref}
\usepackage{graphicx}
\usepackage{bm}
\usepackage{physics}
\usepackage{slashed}
\usepackage{mathtools}  
\usepackage{lipsum}
\crefname{equation}{Eq.}{Eqs.}
\Crefname{equation}{Eq.}{Eqs.}

\title{Boosted energy-energy correlators in the small-acoplanarity limit}

\author[a]{Benjamin Jobilal,}
\author[a,b,c]{Zhong-Bo Kang,}
\author[a,b]{Diego Padilla}

\affiliation[a]{Department of Physics and Astronomy, University of California, Los Angeles, CA 90095, USA}

\affiliation[b]{Mani L. Bhaumik Institute for Theoretical Physics, University of California, Los Angeles, CA 90095, USA}

\affiliation[c]{Center for Frontiers in Nuclear Science, Stony Brook University, NY 11794, USA}

\emailAdd{bjobilal@g.ucla.edu}
\emailAdd{zkang@physics.ucla.edu}
\emailAdd{dpadi022@g.ucla.edu}

\newcommand{\zth}{z_\theta}
\newcommand{\zph}{z_\phi}

\graphicspath{{Images/}}

\abstract{
We study the energy--energy correlator in the hadronic decay of a boosted boson.
We focus on the small-acoplanarity region, where the two hadron momenta are
nearly back-to-back in the plane transverse to the boson momentum.
We derive a transverse-momentum-dependent factorization formula for electroweak
and Higgs bosons in this region, expressed directly in boosted-frame variables
in terms of hard, soft, and jet functions with a one-dimensional recoil measurement.
We verify its singular real-emission terms at next-to-leading order and present
resummed predictions up to N$^3$LL for asymmetric $e^+e^-$ colliders.
For vector bosons, the azimuthal orientation of the hadron pair about the boost
axis is sensitive to the transverse-spin correlation of the produced quarks
through the Collins fragmentation function, which we study for photon-tagged
$Z$ bosons in $e^+e^-\to Z\gamma$ and compare with \textsc{Pythia}~8 supplemented
by \textsc{StringSpinner}.
}

\begin{document}
\maketitle
\section{Introduction} \label{sec:intro}
Energy correlators~\cite{Sterman:1975xv,Basham:1978bw} were among the first infrared-safe observables proposed to test Quantum Chromodynamics (QCD). Of particular interest is the two-point energy correlator, also called the energy-energy correlator (EEC). It measures the energy deposited in two detectors as a function of the angle between them. Sourced by a particle at rest, the EEC displays two interesting behaviors at its endpoints. In the limit that the two detectors are collinear it probes the physics of collinear splittings, with a small-angle evolution governed by splitting-function anomalous dimensions~\cite{Dixon:2019uzg}, while dihadron fragmentation provides an extension to the confinement region~\cite{Kang:2025zto,Lee:2024esz,Chang:2025kgq,Guo:2025zwb}. Experimental measurements have shown that the collinear limit can probe the transition between perturbative and non-perturbative physics in QCD~\cite{CMS:2024mlf,ALICE:2024dfl,STAR:2025jut} and have sparked an abundance of phenomenological studies in this region~\cite{Lee:2022uwt,Barata:2026uby,Barata:2024wsu,Fu:2024pic,Lin:2026knv}. On the other hand, in the limit in which the two detectors are back-to-back the EEC probes the physics of soft and collinear recoil and contains large Sudakov logarithms, whose resummation is naturally organized through TMD factorization~\cite{Moult:2018jzp,Gao:2019ojf,Li:2020bub}. This has enabled numerous phenomenological studies of precision QCD and hadronic and nuclear structure through back-to-back EECs and their generalizations~\cite{Kang:2023big,Bhattacharya:2025bqa,Kang:2024dja,Kang:2024otf,Kang:2023oqj,Barata:2026tis}.

Beyond their intuitive physical interpretation, energy correlators
are theoretically appealing because of their simple analytical
structure and their operator definition in terms of the
energy--momentum tensor (EMT)~\cite{Sveshnikov:1995vi,Tkachov:1995kk,Korchemsky:1999kt}.
This compactness is reflected in their Lorentz transformation
properties: for massless particles, energy-flow operators transform
with a simple, direction-dependent rescaling. This property has
motivated recent phenomenological studies of energy correlators
produced by boosted sources~\cite{Ricci:2022htc,Gao:2026hoj,Holguin:2026vld,Gao:2026xuq}. In the boosted frame, however, back-to-back pairs no longer sit at a fixed opening angle: the boost spreads the Sudakov region of the rest-frame EEC over a range of opening angles, set by the orientation of the pair relative to the boost axis, so the back-to-back factorization of the rest frame cannot simply be recycled at fixed boosted-frame opening angle. In our approach we do not average over all angular dependencies of the EEC in the boosted frame, but rather we study its dependence on the detectors' opening angle and their orientation with respect to the parent boosted boson. In particular, we study the small-acoplanarity limit, $z_\phi=[1+\cos(\phi_1-\phi_2)]/2\ll1$, in which the two detector directions are approximately back-to-back in the transverse plane and hence nearly coplanar with the boson momentum. This variable coincides with the one used for the transverse energy-energy correlator (TEEC) at hadron colliders~\cite{Ali:1984yp,ATLAS:2015yaa}, whose back-to-back limit likewise admits a TMD factorization in which only the recoil perpendicular to the plane of the hard partons is measured~\cite{Gao:2019ojf,Gao:2023ivm,Li:2020bub}. In our case, however, the axis is set by the momentum of the decaying boson rather than by the colliding beams, the detectors are weighted by energy rather than transverse energy, and we keep the opening angle $z_\theta$.

We show that this limit captures the Sudakov region in the boosted frame. By comparing with fixed-order results we identify that in this limit the EEC factorizes into hard, soft and jet functions with a one-dimensional recoil measurement characteristic of TMD factorization~\cite{Boussarie:2023izj}, and we derive the corresponding leading-power factorization formula. Additionally we note that the small-acoplanarity EEC also probes correlations between the spin state of the hard parton pair and the transverse recoil generated during fragmentation. These correlations are encoded by the quark Collins function~\cite{Collins:1992kk} and the TMD fragmentation function for linearly polarized gluons~\cite{Mulders:2000sh}. Depending on the hard process, these contributions can produce azimuthal modulations or modify the recoil distribution. Our observable provides a complementary approach to probing helicity interference with energy correlators~\cite{Chen:2022jhb,Kang:2023big,Chen:2021gdk,Riembau:2025isw,Song:2025bdj,Wang:2026uej,Gao:2026hoj,Ricci:2022htc}.

The paper is organized as follows. In Section~\ref{sec:Decompositions} we define the EEC hadronic tensor with insertions of electroweak and scalar currents and decompose its Lorentz structure such that it can be easily boosted. Then in Section~\ref{sec:BoostFromulas} we derive the boost formula and apply it to the fixed-order EEC formulas reviewed in the previous section. These fixed order results help us to identify that the small-acoplanarity region hints at TMD factorization, which we then derive for the EEC hadronic tensor with insertions of electroweak and scalar currents. In Section~\ref{sec:Phenomenology} we illustrate our small-acoplanarity formalism by presenting predictions for the inclusive EEC in $e^+e^-$ collisions and a charge-tagged EEC in photon-tagged hadronic decay of a $Z$ boson in $e^+e^- \rightarrow Z\gamma$ collisions, where we focus on testing sensitivity to the Collins effect. Finally we conclude in Section~\ref{sec:Conclusions} and present future directions.
\section{Tensor decomposition of the EEC hadronic tensor}
\label{sec:Decompositions}
We define the normalized EEC in $e^+e^-$ collisions (allowing for $\gamma^*/Z$ interference) by 
\begin{equation}
\frac{d\Sigma_{e^+e^-}}{d\Omega_1\,d\Omega_2}
=
\frac{1}{2Q^2\sigma_0 E_{\mathrm{total}}^2}
\sum_{i,j\in\{\gamma,Z\}}
P_i(Q^2)P_j^*(Q^2)\,
L_{\mu\nu}^{ij}
\frac{dW_{ij}^{\mu\nu}}{d\Omega_1\,d\Omega_2}, \label{eq:epem_eec}
\end{equation}
where $\sigma_0$ is the Born-level total cross section for lepton annihilation to hadrons, $E_{\mathrm{total}}$ is the total hadronic energy, $P_i(Q^2)$ are the relevant vector boson propagators with virtuality $Q^2$ (Eq.~\eqref{eq:Propagators}), $L_{\mu\nu}^{ij}$ is the lepton-spin averaged leptonic tensor and $W_{ij}^{\mu\nu}$ is the EEC hadronic tensor. We also define the EEC measured from a hadronically decaying vector boson as
\begin{equation}
\frac{d\Sigma_V}{d\Omega_1\,d\Omega_2}
=
\frac{1}{2M_V\Gamma_{V,0}E_{\mathrm{total}}^2}
\sum_{\lambda,\lambda'}
\rho_{\lambda\lambda'}\,
\frac{dW_{\lambda'\lambda}}{d\Omega_1\,d\Omega_2},
\end{equation}
where now $\Gamma_{V,0}$ is the Born-level decay width of the vector boson to hadrons, $M_V$ is its mass,  $\rho_{\lambda\lambda'}$ is the density matrix of the vector boson and $W_{\lambda'\lambda}$ is the EEC hadronic tensor given by $W_{\lambda'\lambda} = \left(\epsilon^\mu_{\lambda'}\right)^* \,  W_{\mu \nu} \,\epsilon^\nu_{\lambda}$. The analogous EEC sourced by a hadronically decaying scalar particle is obtained by dropping the density matrix and the helicity indices. We now define the EEC hadronic tensor by
\begin{equation}
   \frac{dW_{\mathcal{EE},ij}}{d\Omega_{1}d\Omega_{2}}= \int  d^4x\, e^{i q\cdot x} \langle 0|J^\dagger_i(x)\mathcal{E}_{\mathrm{out}}\left(n_{1}\right)\mathcal{E}_{\mathrm{out}}\left(n_{2}\right) J_j(0)|0 \rangle,
   \label{eq:eec-general-def}
\end{equation}
where the inserted currents create quarks or gluons from the vacuum. The energy-flow operator, with $n^\mu_i = \left(1,\hat{n}_i\right)$ is defined in the COM frame in terms of the EMT as~\cite{Sveshnikov:1995vi,Tkachov:1995kk,Korchemsky:1999kt}
\begin{equation}
     \mathcal{E}_{\mathrm{out}}\left(n\right) =\lim_{r\rightarrow\infty}  \int_0^\infty dt \, r^2 n^i T^{0i}\left(t,r \hat{n}\right) \label{eq:energy_flow_matrix_element_com},
 \end{equation}
where we added the out label to indicate that it acts diagonally on an out-state containing $m$ particles, with $|X;\mathrm{out}\rangle = |k_1,...k_m\rangle$. Indeed its action on such a state is given by
\begin{equation}
\mathcal{E}_\mathrm{out}\left(n\right)|X;\mathrm{out}\rangle = \sum_{i=1}^mk_i^{0}\delta^{2}(\Omega_{n}-\Omega_{k_i})|X;\mathrm{out}\rangle.
\label{eq:energy-flow-fock-def}
\end{equation}
The (out) energy flow operator measures the flux of energy per unit solid angle of the detector at future null (timelike) infinity for massless (massive) particles. This expression also makes clear that the energy-flow operator is not a Lorentz scalar. Under a Lorentz boost, both the particle energies and the angular delta function transform. Its transformation law can be derived from the transformation of the detector directions (see Appendix~\ref{app:BoostDerivation}) and is found to be a simple re-scaling (for massless particles):
\begin{equation}
    U\left(\Lambda\right)\mathcal{E}_{\mathrm{out}}\left(n\right)   U ^\dagger \left(\Lambda\right)= \kappa^{3}\mathcal{E}_{\mathrm{out}}\left(n'\right) \label{eq:BoostEEC},
\end{equation}
where $n'^{\mu}=\kappa\,\Lambda^{\mu}_{\ \nu}n^{\nu}=\left(1,\hat{n}'\right)$ and $\kappa=\left(\Lambda^{0}_{\ \nu}n^{\nu}\right)^{-1}$ (see Eq.~\eqref{eq:n-sphere_boost}). In Section~\ref{sec:BoostFromulas} we will be concerned with boosted energy correlators which can be derived using this formula. Before diving into the derivation it will be useful to inspect the EEC hadronic tensor with insertion of  different sources and decompose its possible structure into scalar projections. In~\cite{Riembau:2025isw} such a decomposition of the EEC hadronic tensor with two vector currents was obtained. Here we will allow for flavor neutral axial and vector currents, as well as flavor changing axial and vector currents.  To constrain the possible Lorentz structure we use the following conditions:

\begin{enumerate}
    \item The energy flow operators commute~\cite{Chicherin:2020azt,Hofman:2008ar}, such that the EEC hadronic tensor is invariant under $n_1\leftrightarrow n_2$.
    \item  For the observables considered here, the EEC hadronic tensor enters
only through contractions with tensors that are transverse to
$q^\mu$ in both Lorentz indices, where $q^\mu$ is the momentum
carried by the current. It is therefore sufficient to retain its
$q$-transverse projection,
\begin{equation}
W_{\perp,ij}^{\mu\nu}
\equiv
g_{\perp,\alpha}^{\mu}\,
g_{\perp,\beta}^{\nu}\,
W_{\mathcal{EE},ij}^{\alpha\beta},
\qquad
g_{\perp}^{\mu\nu}
=
\eta^{\mu\nu}-\frac{q^\mu q^\nu}{Q^2}.
\label{eq:TransverseEECTensorAV}
\end{equation}
For flavor-neutral vector currents, the Ward identity
already guarantees this transversality. By contrast, axial-current
conservation is broken by quark masses and, in the flavor-singlet
case, by the Adler--Bell--Jackiw
anomaly~\cite{Bell:1969ts,Adler:1969gk}, while flavor-changing vector
currents are not conserved for unequal quark masses. 
    \item   The available Lorentz scalars are $z$ and $Q^2$ defined by
    \begin{equation}
    z \equiv \frac{Q^2\left(n_1\cdot n_2\right)}{2\left(q\cdot n_1\right)\left(q\cdot n_2\right)},
    \label{eq:covariant-z-def} \hspace{1cm } q^2=Q^2.
\end{equation}
The available tensor building blocks are $q^\mu$, $\tilde n_1^\mu$,$\tilde n_2^\mu$, $\eta^{\mu\nu}$, and the four-dimensional Levi-Civita tensor $\epsilon^{\mu\nu\alpha\beta}$. Since the normalized detector directions $n_a^\mu=(1,\hat n_a)$ acquire an additional rescaling under Lorentz transformations (see Eq.~\eqref{eq:n-sphere_boost}), we construct Lorentz-covariant vectors by defining
\begin{equation}
    \tilde{n}_a^{\mu} \equiv \frac{Q}{q\cdot n_{a}}\left(n_{a}^{\mu}-\frac{q\cdot n_{a}}{Q^2}q^{\mu}\right),
    \label{eq:ntilde-def}
\end{equation}
which satisfy $\tilde{n}_a \cdot \tilde{n}_a=-1$, $\tilde{n}_1 \cdot \tilde{n}_2=2 z-1$ and $q\cdot \tilde{n}_a=0$ with $a=1,2$. 
\end{enumerate}

\subsection{EEC with axial and vector flavor-neutral currents }
In this subsection we begin by studying the EEC hadronic tensor measured on states created by flavor-neutral axial and vector currents. It reads
\begin{equation}
   \frac{dW^{\mu\nu}_{\mathcal{EE},ij}}{d\Omega_{1}d\Omega_{2}}= \int  d^4x\, e^{i q\cdot x} \langle 0|J^{\mu\dagger}_i(x)\mathcal{E}_{\mathrm{out}}\left(n_{1}\right)\mathcal{E}_{\mathrm{out}}\left(n_{2}\right) J^{\nu}_j(0)|0 \rangle, \hspace{1 cm} i = {A,V}.
   \label{eq:eec-tensor-def}
\end{equation}Here we use the notation

\begin{equation}
   J_V^\mu\left(x\right) = \bar{q}^a\left(x\right)\gamma^\mu q^a\left(x\right),\hspace{0.7 cm}  J_A^\mu\left(x\right) = \bar{q}^a\left(x\right)\gamma^\mu\gamma^5 q^a\left(x\right)
    \label{eq:VA_currents},
\end{equation}
for the coupling-stripped vector and axial currents and $q\left(x\right)$ are quark fields with summed-over color index $a$.  The only difference from the vector-current-only decomposition in \cite{Riembau:2025isw} could arise from the axial-vector interference in $ W_{\perp,VA}^{\mu\nu}, W_{\perp,AV}^{\mu\nu}$ (the double axial term shares the same decomposition as the double vector one since $J_AJ_A$ obeys the same discrete transformation properties as $J_VJ_V$). However, since the energy flow operator is charge blind, by applying charge conjugation in Eq.~\eqref{eq:eec-tensor-def}  we immediately find that these terms vanish
\begin{align}
W^{\mu\nu}_{\perp,VA}&=W^{\mu\nu}_{\perp,AV}=0.
\end{align}
Therefore the decomposition of Eq.~\eqref{eq:TransverseEECTensorAV} is given by~\cite{Riembau:2025isw}  as
\begin{equation}
W^{\mu\nu}_{\perp,i}=-\frac{g_\perp^{\mu\nu}}{3}\,W_{0,i}(z,Q^2)
+\Big(C^{\mu\nu}+\frac{g_\perp^{\mu\nu}}{3}\Big)W_{1,i}(z,Q^2)
+\Big(D^{\mu\nu}+\frac{g_\perp^{\mu\nu}}{3}\Big)W_{2,i}(z,Q^2)\label{eq:VA_decomposition}
\end{equation}where the index $i=VV,AA$ indicates whether this is a tensor constructed with two vector or two axial currents. The projectors are given in terms of 
\begin{equation}
    e_c^\mu \equiv
    \frac{(\tilde{n}_1+\tilde{n}_2)^\mu}{2\sqrt{1-z}},
    \qquad
    e_d^\mu \equiv
    \frac{(\tilde{n}_1-\tilde{n}_2)^\mu}{2\sqrt{z}},
    \label{eq:detector-directions}
\end{equation}which satisfy $q\cdot e_a=0$ and
$e_a\cdot e_b=-\delta_{ab}$, with $a,b\in\{c,d\}$.
The tensor structures can then be written as
\begin{align}
    C^{\mu\nu} &= e_c^\mu e_c^\nu,
    &
    D^{\mu\nu} &= e_d^\mu e_d^\nu,
    \label{eq:C-projector-def}
\end{align}
and the scalar projections are related to Eq.~\eqref{eq:TransverseEECTensorAV} by
\begin{align}
    W_{0,i}(z) &=- g_{\perp\mu\nu}\, W^{\mu\nu}_{\perp,i}, \label{eq:H0proj}\\[4pt]
  W_{1,i}(z) &= \big( 2\, C_{\mu\nu} + D_{\mu\nu} + g_{\perp\mu\nu} \big)\,
                W^{\mu\nu}_{\perp,i}, \label{eq:H1proj}\\[4pt]
  W_{2,i}(z) &= \big( C_{\mu\nu} + 2\, D_{\mu\nu} +g_{\perp\mu\nu} \big)
                 W^{\mu\nu}_{\perp,i}. \label{eq:H2proj}
\end{align}
In the rest frame of the vector boson and for massless quarks, the three projections at $\mathcal{O}(\alpha_s)$ are the same for two vector and for two axial-vector currents:
\begin{align}
\frac{d W_{0,i}(z,Q^2)}{d\Omega_1d\Omega_2} &= \frac{\alpha_s C_F N_c\, Q^4}{128\pi^4}\,
\frac{3-2z}{(1-z)\,z^5}
\Big[\, 3z(2-3z) + 2\left(3-6z+2z^2\right)\log(1-z) \Big],\label{eq:porjectionsCOM0} \\
\frac{d W_{1,i}(z,Q^2)}{d\Omega_1d\Omega_2} &= -\frac{\alpha_s C_F N_c\, Q^4}{256\pi^4\, z^5}
\Big[\, z\left(18-15z-2z^2\right) + 2\left(9-12z+2z^2\right)\log(1-z) \Big]\label{eq:porjectionsCOM1}, \\
\frac{d W_{2,i}(z,Q^2)}{d\Omega_1d\Omega_2} &= \frac{\alpha_s C_F N_c\, Q^4}{256\pi^4\,(1-z)\, z^4}
\Big[\, z\left(6-5z+2z^2\right) + 2\left(3-4z+2z^2\right)\log(1-z) \Big] \label{eq:porjectionsCOM2}.
\end{align}
In the literature the EEC in $e^+e^-$ annihilation is usually discussed in the center of mass frame with the electron directions integrated over, such that only ${W}_0(z)$ is relevant~\cite{Riembau:2025isw,Chicherin:2020azt}. Note that the 4-dimensional Levi-Civita symbol did not appear in the decomposition; it will only contribute to energy-correlators sourced by parity-violating currents.
\subsection{EEC with axial and vector flavor-changing currents}

We now study the case of flavor changing currents such as those coupling the $W^+$ and $W^-$ to left-handed quarks. This EEC is more interesting since it allows for a non-zero contribution from axial-vector mixing. The two relevant coupling-stripped currents are given by
\begin{equation}
    J_+^\mu\left(x\right) = \, \bar{u}^a\left(x\right)\gamma^\mu \frac{1-\gamma^5}{2}d^a\left(x\right),\hspace{0.7 cm}   J_-^\mu\left(x\right) =  \left( J_+^\mu\left(x\right) \right)^\dagger,
    \label{eq:Ws_currents}
\end{equation}
where $u\left(x\right)$ represents a quark field of flavor $u \, \mathrm{or  }\, c$ and $d\left(x\right)$ represents a quark field of flavor $d,s\, \mathrm{or }\, b$. We will again obtain the tensor decomposition of the $q-$transverse part of the  EEC-hadronic tensor. For two currents coupling a $W^+$ to left-handed quarks we will use the notation $W_{\perp,++}^{\mu\nu}$. In contrast to the previous subsection, charge conjugation does not force the vector--axial interference contribution to vanish. Instead, it relates the corresponding interference terms in
$W^+$ and $W^-$ decays. To see this let us decompose the chiral current into vector and axial pieces by introducing the notation $V_+=\bar u_i\gamma^\mu d_j$ and $A_+=\bar u_i\gamma^\mu\gamma^5 d_j$,
$V_-=(V_+)^\dagger$ and $A_-=(A_+)^\dagger$. It is then easy to check by charge conjugation that
\begin{align}
W^{\mu\nu}_{\perp,V_+V_+}&=W^{\mu\nu}_{\perp,V_-V_-}\,, & \label{eq:C-ConjugationVV}
W^{\mu\nu}_{\perp,A_+A_+}&=W^{\mu\nu}_{\perp,A_-A_-}\,,\\
W^{\mu\nu}_{\perp,V_+A_+}&=-W^{\mu\nu}_{\perp,V_-A_-}\,\label{eq:C-Conjugation}, &
W^{\mu\nu}_{\perp,A_+V_+}&=-W^{\mu\nu}_{\perp,A_-V_-}\,.
\end{align}
However observe that when considering a purely QCD final state (thus neglecting weak interactions in the final state which could spoil flavor symmetry) the relevant scattering amplitudes are completely symmetric up to $m_u \leftrightarrow m_d$, from which it follows that
\begin{equation}
    W^{\mu\nu}_{\perp,V_+A_+}\left(m_{u},m_{d}\right) =  -W^{\mu\nu}_{\perp,V_+A_+}\left(m_{d},m_{u}\right),
\end{equation}
and the same expression holds for the reversed order, $W^{\mu\nu}_{\perp,A_+V_+}$. Then the vector-axial interference only gives a non-zero contribution to the EEC if we consider quarks of unequal masses. We now use parity  conjugation on the operator definition (see Eq.~\eqref{eq:parity_EMT} for a discussion on the transformation of the energy flow operator under parity), which yields the following relations (we suppress the $\pm$ indices since parity is flavor diagonal)
\begin{equation}
    W_{\perp,VV}^{\mu\nu}\left(n_1,n_2,q\right) =P^{\mu}_{\alpha}P^{\nu}_{\beta} W_{\perp,VV}^{\alpha\beta}\left(P\cdot n_1,P\cdot n_2,P\cdot q\right),
\end{equation}\begin{equation}
    W_{\perp,AA}^{\mu\nu}\left(n_1,n_2,q\right) =P^{\mu}_{\alpha}P^{\nu}_{\beta} W_{\perp,AA}^{\alpha\beta}\left(P\cdot n_1,P\cdot n_2,P\cdot q\right),
\end{equation}\begin{equation}
    W_{\perp,VA}^{\mu\nu}\left(n_1,n_2,q\right) =-P^{\mu}_{\alpha}P^{\nu}_{\beta} W_{\perp,VA}^{\alpha\beta}\left(P\cdot n_1,P\cdot n_2,P\cdot q\right),\label{eq:VA-Parity}
\end{equation}\begin{equation}
    W_{\perp,AV}^{\mu\nu}\left(n_1,n_2,q\right) =-P^{\mu}_{\alpha}P^{\nu}_{\beta} W_{\perp,AV}^{\alpha\beta}\left(P\cdot n_1,P\cdot n_2,P\cdot q\right), \label{eq:AV-Parity}
\end{equation}where $P_{\nu}^\mu=\mathrm{diag}\left(1,-1,-1,-1\right)$. As claimed earlier, to satisfy Eqs.~\eqref{eq:AV-Parity}  and~\eqref{eq:VA-Parity} which are due to the parity-violating vector-axial mixing, it will be necessary to add the 4-dimensional Levi-Civita symbol into the available Lorentz structures. To find the most general decomposition we therefore introduce the normal direction
\begin{equation}
    e_N^\mu \equiv
    \epsilon^{\mu\alpha\beta\gamma}
    \frac{q_\alpha}{Q}e_{d\beta}e_{c\gamma},
    \label{eq:normal-direction}
\end{equation}
in terms of which the most general decomposition of the $W^+$ decay EEC is found to be
\begin{equation}
\begin{split}
W^{\mu\nu}_{\perp,++}
={}&-\frac{g_\perp^{\mu\nu}}{3}W_{0,++}(z,Q^2)
+\Big(C^{\mu\nu}+\frac{g_\perp^{\mu\nu}}{3}\Big)W_{1,++}(z,Q^2)\\
&+\Big(D^{\mu\nu}+\frac{g_\perp^{\mu\nu}}{3}\Big)W_{2,++}(z,Q^2)
+A^{\mu\nu}\,W_{A,++}(z)+S^{\mu\nu}\,W_{S,++}(z,Q^2)\,. \label{eq:W-EEC-decomp}
\end{split}
\end{equation}
The first three terms in Eq.~\eqref{eq:W-EEC-decomp} have the same decomposition as Eq.~\eqref{eq:VA_decomposition} and the two new structures are given by 
\begin{align}
    A^{\mu\nu}
    &= i\left(
        e_N^\mu e_d^\nu-e_d^\mu e_N^\nu
    \right),
    \label{eq:A-projector-def}
    \\
    S^{\mu\nu}
    &= e_N^\mu e_d^\nu+e_d^\mu e_N^\nu,
    \label{eq:S-projector-def}
\end{align}
with the corresponding scalars obtained from
\begin{align}
  W_{A,++}(z,Q^2) &= \frac{1}{2}\, A^{*}_{\mu\nu}\,
                 W^{\mu\nu}_{\perp,++}, \label{eq:HAproj}\\[4pt]
  W_{S,++}(z,Q^2) &= \frac{1}{2}\, S_{\mu\nu}\,
                 W^{\mu\nu}_{\perp,++}. \label{eq:HSproj}
\end{align}
The structure $A^{\mu\nu}$ is odd under $\mu\leftrightarrow\nu$. The factor of $i$ in its definition is required by the hermiticity of the hadronic tensor,
\begin{equation}
    W^{\mu\nu}_{\perp,++} = \left(W_{\perp,++}^{\nu\mu}\right)^*. \label{eq:Hermiticity-WEEC}
\end{equation}
Since charge and parity conjugation were instructive to find this decomposition, we naturally ask whether anything can be learned from studying time reversal conjugation of the energy flow operator. To investigate this let us first recall that the energy flow operator inserted in the EEC hadronic tensor acts diagonally on out-states (see Eq.~\eqref{eq:energy-flow-fock-def}). However, conjugating this energy flow operator with time-reversal we get $\mathcal{T}\mathcal{E}_{\mathrm{out}}\left(n\right)\mathcal{T}^{-1}  = \mathcal{E}_{\mathrm{in}}\left(Pn\right)$ (see Eq.~\eqref{eq:T-reversal_conjugationEEC}), where now $\mathcal{E}_{\mathrm{in}}\left(n\right)$ acts diagonally on in-states (see Eq.~\eqref{eq:Eflow_in}). Acting with time-reversal on the full EEC hadronic tensor, one finds
\begin{equation}
W^{\mu\nu}_{\mathcal{E}\mathcal{E},++}(n_1,n_2,q)
= P^{\mu}_{\alpha}P^{\nu}_{\beta}
\left(\int d^4x\, e^{i(P q)\cdot x}\,
\langle 0|\, J_+^{\alpha\dagger}(x)\,
\mathcal{E}_{\mathrm{in}}(P\cdot n_1)\,
\mathcal{E}_{\mathrm{in}}(P\cdot n_2)\,
J_+^{\beta}(0)\,|0\rangle\right)^{*} .
\label{eq:T_EEC}
\end{equation}
Projecting Eq.~\eqref{eq:T_EEC} onto the $q$-transverse subspace commutes with time reversal, since $g_\perp^{\mu\nu}$ is real, symmetric and satisfies $P^{\mu}_{\alpha}P^{\nu}_{\beta}\,g_\perp^{\alpha\beta}(P\cdot q)=g_\perp^{\mu\nu}(q)$. Contracting the result with the projectors of Eq.~\eqref{eq:W-EEC-decomp} and using the hermiticity relation~\eqref{eq:Hermiticity-WEEC}, which holds equally for in-detectors, we find that each scalar function satisfies
\begin{equation}
W_{a,++}(z,Q^2)\Big|_{\mathrm{out}}=\pi_a\,\rho_a\,W_{a,++}(z,Q^2)\Big|_{\mathrm{in}}\,,
\qquad a=0,1,2,A,S\,,
\label{eq:T_scalars}
\end{equation}
where the correlators on the LHS have out-detector insertions and those on the RHS have in-detector insertions, $\rho_a=+1$ ($-1$) if the corresponding projector is symmetric (antisymmetric) under $\mu\leftrightarrow\nu$, and $\pi_a=-1$ if it contains a Levi-Civita tensor ($a=A,S$) and $\pi_a=+1$ otherwise. Hence $W_0$, $W_1$, $W_2$ and $W_A$ take the same values whether they are measured with out- or in-energy flow operators, while $W_S$ changes sign. Equivalently,
\begin{equation}
W^{\mu\nu}_{\perp,++}\Big|_{\mathrm{out}}-W^{\mu\nu}_{\perp,++}\Big|_{\mathrm{in}}
=2\,S^{\mu\nu}\,W_{S,++}(z,Q^2)\,,
\label{eq:Delta_Hperp}
\end{equation}
that is,
\begin{equation}
W_{S,++}(z,Q^2)=\frac{1}{4}\,S_{\mu\nu}\int d^4x\,e^{iq\cdot x}\,
\langle 0|\,J_+^{\mu\dagger}(x)\Big[\mathcal{E}_{\mathrm{out}}(n_1)\,\mathcal{E}_{\mathrm{out}}(n_2)
-\mathcal{E}_{\mathrm{in}}(n_1)\,\mathcal{E}_{\mathrm{in}}(n_2)\Big]J_+^{\nu}(0)\,|0\rangle\,,
\label{eq:HS_inout}
\end{equation}
where the $g_\perp$ projectors have been dropped since $S^{\mu\nu}$ is itself $q$-transverse. $W_{S,++}(z,Q^2)$ is therefore a naive time-reversal-odd quantity. In perturbation theory it comes from absorptive parts of loop diagrams, which cancel for massless quarks at $\mathcal{O}(\alpha_s^2)$~\cite{Brandenburg:1995nv,Korner:2000zr} and are suppressed by $m_c^2/M_W^2$ for $W \rightarrow c\bar{s}$, so that non-perturbative phases generated during hadronization could provide the dominant contribution. We leave a quantitative study of this possibility to future work. The hadronic tensor $W^{\mu\nu}_{\mathcal{E}\mathcal{E},++}$ defined either with in or out detectors admits a simple physical interpretation. For the case with insertions of $\mathcal{E}_{\mathrm{out}}$, the current creates a $W^{+}$ at rest at the origin, which decays into quarks and gluons. These hadronize, and reach a calorimeter at future timelike infinity (since this is non-vanishing only for massive quarks), where the energy flowing along $n_1$ and $n_2$ is recorded. The same tensor with $\mathcal{E}_{\mathrm{out}}\to\mathcal{E}_{\mathrm{in}}$ describes the time-reversed (formation) process: incoming hadrons, whose energy flow along their directions of motion $n_1$ and $n_2$ is recorded at past timelike infinity, converge on the origin and fuse into a $W^{+}$. Energy conservation guarantees that the total energy recorded is the same in both pictures, $\int d\Omega\,\mathcal{E}_{\mathrm{in}}(n)=\int d\Omega\,\mathcal{E}_{\mathrm{out}}(n)=P^0$, but the angular correlations need not coincide, since the particles interact after their production in one case and before their annihilation in the other. Time-reversal invariance strongly restricts this difference: for the $q$-transverse tensor it is carried entirely by $W_S$. Let us now look at the diagrams that contribute to Eq.~\eqref{eq:HS_inout}. Introducing complete sets of out-states, identifying the overlap between in-states and out-states with the S-matrix $S_{XY}=\langle X;\mathrm{out}|Y;\mathrm{in}\rangle =\langle X;\mathrm{out}|S|Y;\mathrm{out}\rangle $ \cite{Caron-Huot:2023vxl}  and defining the out-state amplitudes $\mathcal{M}_{X}^\mu\equiv \langle X;\mathrm{out}|J^\mu\left(0\right)|0\rangle $, after some algebra the matrix element in Eq.~\eqref{eq:HS_inout} can be rewritten as
\begin{align}
&\int d^4x\, e^{iq\cdot x}\,\langle 0|\, J_+^{\mu\dagger}(x)
\Big[\mathcal{E}_{\mathrm{out}}(n_1)\,\mathcal{E}_{\mathrm{out}}(n_2)
-\mathcal{E}_{\mathrm{in}}(n_1)\,\mathcal{E}_{\mathrm{in}}(n_2)\Big]
J_+^{\nu}(0)\,|0\rangle \nonumber\\
&\qquad=\sum_X (2\pi)^4\delta^4(q-p_X)
\sum_{i,j=1}^{m_X} k_i^0\, k_j^0\,
\delta^2(\Omega_{n_1}-\Omega_{k_i})\,\delta^2(\Omega_{n_2}-\Omega_{k_j}) \nonumber\\
&\qquad\quad\times\Big[-i\sum_Y \big(\mathcal{M}^\mu_Y\big)^* T_{YX}\,\mathcal{M}^\nu_X
+i\sum_Y \big(\mathcal{M}^\mu_X\big)^* T^\dagger_{XY}\,\mathcal{M}^\nu_Y
-\sum_{Y,Z}\big(\mathcal{M}^\mu_Y\big)^* T_{YX}\,T^\dagger_{XZ}\,\mathcal{M}^\nu_Z\Big],\label{eq:HS_diagrams}
\end{align}
where we used $S=1+iT$ such that $T_{YX} = \langle Y;\mathrm{out}|T|X;\mathrm{out}\rangle$. The sums over final states can be interpreted as cut diagrams, which indeed shows that $W_S$ is an interference effect coming from the imaginary part of loop diagrams. The first two terms in Eq.~\eqref{eq:HS_diagrams} are related by complex conjugation and swap of Lorentz indices, and therefore it suffices to compute one of them, their first non-trivial instance starting at $\mathcal{O}\left(\alpha_s^2\right)$. The first non-trivial contribution to the last term in Eq.~\eqref{eq:HS_diagrams} only starts at $\mathcal{O}\left(\alpha_s^3\right)$. Note that when the states $X$ and $Y$ contain more than two particles, $T_{YX}$ also contains disconnected (also called spectator) contributions, in which only a subset of the particles rescatter~\cite{Jackura:2018xnx}, and these must be kept when evaluating the cut diagrams of Eq.~\eqref{eq:HS_diagrams}.

Finally note that as a result of the charge and parity constraints discussed above the EEC hadronic tensor for hadronic decay of a $W^-$ is related to that of a $W^+$ by:
\begin{equation}
    W_{a,--}=W_{a,++}\hspace{0.6 cm} (a=0,1,2),\hspace{1cm }W_{a,--}=-W_{a,++}\hspace{0.6 cm} (a=A,S).
\end{equation}
\subsection{EEC with scalar sources}
\label{sec:Hdecay}
For an energy correlator coupled to scalar sources we will consider a Higgs boson undergoing hadronic decay. The two relevant currents are given by
\begin{equation}
    J_{Y}\left(x\right) = \bar{q}^a\left(x\right)q^a\left(x\right),\hspace{1.5 cm} J_{G}\left(x\right) = G^{b,\mu\nu}\left(x\right)G^b_{\mu\nu}\left(x\right).
\end{equation}
 $G^{b,\mu\nu}\left(x\right)$ is the gluon field strength tensor with adjoint color index $b$ which is summed over in the current. Using these currents in Eq.~\eqref{eq:eec-general-def} four different EEC hadronic Higgs decay correlators can be formed. However it is common practice to keep a non-zero Yukawa coupling, while at the same time neglecting the mass of the quarks. In this setting, the mixed terms $(W_{YG},W_{GY})$ vanish since the Yukawa current $J_Y$ mixes chiralities while the quarks coming from QCD interactions in $J_G$ do not. A mixed term would only appear for non-zero quark masses, which is also when chirality is not a good quantum number~\cite{Bi:2020frc,Gao:2019mlt}.

In the center-of-mass frame, with massless quarks the two possible EECs are given by~\cite{Gao:2020vyx,Luo:2019nig}
\begin{equation}
    \tilde{W}_{YY}  = -\frac{\alpha_s C_FN_c Q^4}{256\pi^4}\frac{3z\left(6-5z\right)+2\left(z^3+3z^2-12z+9\right)\ln{\left(1-z\right)}}{\left(1-z\right)z^5}, \label{eq:HiggsQuarks}
\end{equation}
\begin{align}
  \tilde{W}_{GG}
    &= \frac{\alpha_s\,(N_c^2-1)\,Q^6}{64\pi^4}
       \Big[ C_A\, f_{lc}(z) + T_F\, n_f\, f_{n_f}(z) \Big],\label{eq:HiggsGluons} \\
  f_{lc}
    &= \frac{z\left(25z^3 - 156z^2 + 336z - 216\right)
             - 6\left(2z^4 - 14z^3 + 51z^2 - 74z + 36\right)\ln(1-z)}
            {6z^6(1-z)}, \\
  f_{n_f}
    &= \frac{6\left(z^4 - 17z^3 + 63z^2 - 83z + 36\right)\ln(1-z)
             - z\left(25z^3 - 201z^2 + 390z - 216\right)}
            {3z^6(1-z)}.
\end{align}
The tildes in the equations above indicate that the overall coupling constants have been stripped.
\section{Boosted EECs and factorization at small acoplanarity}
\label{sec:BoostFromulas}
\begin{figure}[t]
  \centering
  \includegraphics[width=0.65\textwidth]{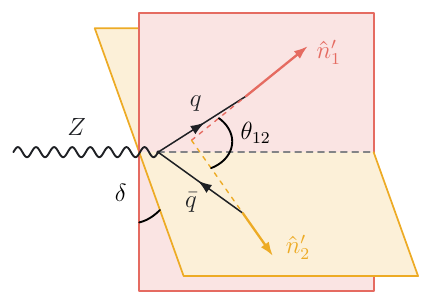}\\[2pt]
  \makebox[0.15\textwidth]{}\makebox[0.15\textwidth]{}
  \caption{Illustration of the EEC sourced by a boosted $Z$ boson. Here $\theta_{12}$ is the angle between the two detectors (detected hadrons) and $\delta $ is the angle between the $\hat{n}_1$-$Z$ and the $\hat{n}_2$-$Z$ planes. As the detected hadrons and the $Z$ boson become coplanar $z_\phi$ approaches zero since for small $\delta$ we can approximate $z_\phi\approx\delta^2/4$. }
  \label{fig:EEC_Boosted}
\end{figure}

Having decomposed the Lorentz structure of the EEC hadronic tensor, we can boost each of the projections $W_{a}(z,Q^2)$ with the boost formula in Eq.~\eqref{eq:BoostEEC}. Since this transformation only holds for massless quarks, in what follows we will neglect quark masses, which removes $W_{A}$ and $W_{S}$ from the $W^{\pm}$ EEC.  As reviewed above, in the center-of-mass frame the projections only depend non-trivially on the Lorentz scalar $z$ which is the opening angle between pairs of correlated final-state particles. Once we boost to a frame in which the parent boson carries nonzero spatial momentum,
\begin{equation}
    q^\mu = Q\left(\cosh\eta,\,0,\,0,\,\sinh\eta\right)\,, \label{eq:boostedq}
\end{equation}
with $\eta$ the boson rapidity, the opening angle between pairs is aberrated and the boost axis $\hat{z}$ provides a preferred direction with respect to which azimuthal correlations can be measured. To keep track of these modulations let us introduce the following notation for the opening angle between pairs of particles in the boosted frame by
\begin{equation}
z_\theta
=
\frac{1-\hat n_1\cdot\hat n_2}{2},
\end{equation}
as well as the acoplanarity variable
\begin{equation}
z_\phi
=
\frac{1+\cos(\phi_1-\phi_2)}{2} \label{eq:zphi_def}\, ,
\end{equation}
which is insensitive to the boost. The acoplanarity variable is particularly interesting since as it is tuned to zero the parent boson and two detected hadrons become approximately coplanar, as illustrated in Fig.~\ref{fig:EEC_Boosted} for a $Z$ boson which decays into two hadrons. 
Using Eq.~\eqref{eq:BoostEEC} we find a formula for the EEC boosted from the center-of-mass frame to the frame defined by the boson momentum in Eq.~\eqref{eq:boostedq}
\begin{equation}
\begin{aligned}
  \frac{dW_{a}(z_\theta,z_{\phi};\eta)}{d z_\theta dz_{\phi}} &=  \int d\Omega'_1d\Omega'_2 \      \frac{dW_{a}\left(z\right)}{d \Omega'_1d\Omega'_2} \delta\left(z_\theta-\frac{1-\hat{n}_1\cdot\hat{n}_2}{2}\right)\delta\left(z_{\phi}-\frac{1+\cos(\phi _1-\phi_2)}{2}\right) \\ &=4\pi \int_{-\infty}^{\infty} d\tilde{\eta} \ \      \frac{dW_{a}\left(z\right)}{d \Omega_1d\Omega_2}\left(\frac{z}{z_\theta}\right)^3\mathcal{J}\left(z_{\theta},z_{\phi},\tilde{\eta}\right)\Theta\left(z_{\theta}-(1-z_{\phi})\sech^2{\tilde{\eta}}\right) ,
    \label{eq:boosted-fixed-order-integral}
\end{aligned}
\end{equation}
where the variable z is now a function of the variables in the boosted frame
\begin{equation}
    z(z_{\theta},z_{\phi},\tilde{\eta};\eta) = z_\theta \frac{z_\phi+\sinh^2\tilde{\eta}}{z_\phi+z_\theta \sinh^2\tilde{\eta}+\left(1-z_\theta\right)\sinh^2{\left(\tilde{\eta}-\eta\right)}}
    \label{eq:z-boosted-variables}
\end{equation}
which satisfies $z(z_{\theta},z_{\phi},\tilde{\eta};0)=z_{\theta}$ as expected. In writing these expressions we introduced the average rapidity of the correlated pair
\begin{equation}
     \tilde{\eta} = \frac{\eta_1+\eta_2}{2},
\end{equation}
and the Jacobian factor $\mathcal{J}\left(z_{\theta},z_{\phi},\tilde{\eta}\right)$ is given by
\begin{equation}
    \mathcal{J}\left(z_{\theta},z_{\phi},\tilde{\eta}\right) = \frac{2\left(1-z_{\theta}\right)}{\left(\sinh^2{\tilde{\eta}+z_{\phi}}\right)\sqrt{z_{\phi}\left(1-z_{\phi}\right)\left(z_{\theta}\sinh^2{\tilde{\eta}}+z_{\phi}\right)\left(z_{\theta}\sinh^2{\tilde{\eta}+z_{\phi}-(1-z_\theta)}\right)}}
    \label{eq:angular-jacobian}.
\end{equation}
The usual one-dimensional EEC is recovered by integrating over the azimuthal correlations,
\begin{equation}
\frac{dW_a}{dz_\theta}
=
\int_0^1 dz_\phi\,
\frac{dW_a}{dz_\theta\,dz_\phi}\label{eq:1D_EED}.
\end{equation}
It was shown in~\cite{Gao:2026xuq} that the acoplanarity-averaged boost kernel in Eq.~\eqref{eq:1D_EED} can be rewritten in terms of a complete elliptic integral of the first kind. Its generalization for boosted polarized decays was recently worked out in~\cite{Gao:2026hoj}.

A natural question is how the back-to-back endpoint enhancement, located at $z\rightarrow1$, manifests itself in the boosted-frame variables. This question is central to our analysis, since its answer determines how the back-to-back factorization of the EEC should be formulated in terms of $z_\theta$ and $z_\phi$. Because the boost does not act on the two particles in the same way, back-to-back pairs no longer sit at a fixed opening angle: a pair that is back-to-back in the rest frame of the boson is folded toward the boost direction, and its opening angle in the boosted frame depends on the orientation of the pair relative to the boost axis. The boost therefore spreads the Sudakov region of the rest-frame EEC over a range of $z_\theta$. What it preserves is coplanarity, since a back-to-back pair and the boost axis always lie in a common plane. As a result, the back-to-back endpoint enhancement is mapped to the region $z_\phi \sim \left(\tilde{\eta}-\eta\right)^2 \ll 1$ (Eq.~\eqref{eq:boosted_hint}), that is, to the limit of small acoplanarity. Applied to Eq.~\eqref{eq:boosted-fixed-order-integral} this indicates that the leading $z_\phi \rightarrow0$ asymptotic behavior of the two-dimensional EEC is obtained precisely by integrating over the $z_\phi \sim \left(\tilde{\eta}-\eta\right)^2 \ll 1$ region. This calculation is carried out explicitly in Appendix~\ref{app:Derivation_zphi_fixedorder}. For energy correlators sourced by $\{\gamma^*,Z,W^{\pm}\}$ coupled to massless quarks, keeping the tree-level contribution and the $\mathcal{O}(\alpha_s)$ real emission at $z_\phi \ll1$, the boosted-frame projections take the form
\begin{subequations}\label{eq:endpoint}
\begin{align}
\frac{dW_0(z_\theta,z_\phi;\eta)}{dz_\theta\,dz_\phi} &= \phantom{-}2\,\mathcal{G}_q(z_\theta,z_\phi;\eta) + \mathcal{O}\big(\alpha_s z_\phi^{-1/2}\big)\,, \label{eq:endpoint_H0}\\[4pt]
\frac{dW_1(z_\theta,z_\phi;\eta)}{dz_\theta\,dz_\phi} &= \mathcal{O}\big(\alpha_s z_\phi^{-1/2}\big)\,, \label{eq:endpoint_H1}\\[4pt]
\frac{dW_2(z_\theta,z_\phi;\eta)}{dz_\theta\,dz_\phi} &= -\,\mathcal{G}_q(z_\theta,z_\phi;\eta) + \mathcal{O}\big(\alpha_s z_\phi^{-1/2}\big)\,, \label{eq:endpoint_H2}
\end{align}
\end{subequations}
where
\begin{equation}
\mathcal{G}_q(z_\theta,z_\phi;\eta) = \frac{N_c\,Q^4}{16\pi}\,\mathcal{K}(z_\theta;\eta)
\left\{\,\delta(z_\phi) - \frac{\alpha_s C_F}{2\pi}\,\frac{1}{z_\phi}
\left[3+2\log\!\left(\frac{4z_\phi(1-z_\theta)}{z_\theta\sinh^2\eta}\right)\right]\right\},
\label{eq:endpoint-Gq}
\end{equation}
with $\mathcal{K}(z_\theta;\eta)$ given by
\begin{equation}
\mathcal{K}(z_\theta;\eta) = \frac{|\csch\eta|\;\Theta\big(z_\theta-\sech^2\eta\big)}{z_\theta^{5/2}\sqrt{z_\theta\cosh^2\eta-1}}\,.
\label{eq:endpoint-K}
\end{equation}
For Higgs decay to quarks (Eq.~\eqref{eq:HiggsQuarks}), the boosted distribution has a similar form
\begin{equation}
\frac{d\tilde{W}_{YY}(z_\theta,z_\phi;\eta)}{dz_\theta\,dz_\phi} = \,\mathcal{G}_q(z_\theta,z_\phi;\eta) + \mathcal{O}\big(\alpha_s z_\phi^{-1/2}\big)\,, \label{eq:endpoint_Hqq}
\end{equation}
while the boosted version of the Higgs decay to gluons EEC (Eq.~\eqref{eq:HiggsGluons}) takes the form
\begin{equation}
    \frac{d\tilde{W}_{GG}}{dz_\theta dz_\phi} = \mathcal{G}_g(z_\theta,z_\phi;\eta)+ \mathcal{O}\big(\alpha_s z_\phi^{-1/2}\big)\,,  \label{eq:HiggsGluonFO}
\end{equation}
where
\begin{equation}
\mathcal{G}_g(z_\theta,z_\phi;\eta) = \frac{\left(N_c^2-1\right)\,Q^6}{8\pi}\,\mathcal{K}(z_\theta;\eta)
\left\{\,\delta(z_\phi) - \frac{\alpha_s }{2\pi}\,\frac{1}{z_\phi}
\left[\beta_0+2C_A\log\!\left(\frac{4z_\phi(1-z_\theta)}{z_\theta\sinh^2\eta}\right)\right]\right\},
\label{eq:endpoint-Gg}
\end{equation}
and $\beta_0 = 1/3\left(11C_A -4T_F n_f\right)$ is the leading-order coefficient of the QCD beta function. Note that the endpoint singularities in $z_\theta$ are integrable over the physical range $\sech^2\eta<z_\theta<1$. Sub-leading corrections in $z_{\phi}$ are also integrable, but the leading $\mathcal{O}(z_{\phi}^{-1})$ and $\mathcal{O}(z_{\phi}^{-1}\log{z_{\phi}})$ reflect the typical infrared enhancement present in the back-to-back limit of the center of mass EEC. Nevertheless, in these two variables ($z_\theta,z_\phi$) the endpoint divergence only affects the azimuthal correlation. We therefore expect a TMD factorization formula in $z_\phi$~\cite{Boussarie:2023izj,Moult:2018jzp,Gao:2023ivm}, valid for $l_\phi\ll p_T$, where $l_\phi$ is the recoil momentum associated with $z_\phi$ (Eq.~\eqref{eq:phi_LP}). The recoil is perturbative for $l_\phi\gg\Lambda_{\mathrm{QCD}}$, and for $l_\phi\sim\Lambda_{\mathrm{QCD}}$ the TMD functions require non-perturbative input. Let us set the kinematics where such a factorization naturally appears.
Consider the Born-level two-body kinematics of two quarks (or gluons) produced from a hard interaction with a boosted boson. Their momenta are given by
\begin{equation}
    \begin{aligned}
    p_{1}^{\mu}&=p_{T}\left(\cosh{\eta_{1},\cos{\phi_1},\sin{\phi_1},\sinh{\eta_{1}}} \right), \\ 
    p_{2}^{\mu}&=p_{T}\left(\cosh{\eta_{2},-\cos{\phi_1},-\sin{\phi_1},\sinh{\eta_{2}}} \right)
    \end{aligned}
    \label{eq:boosted-born-momenta}
\end{equation}
which satisfy $q=p_1+p_2$. Here q, defined in Eq.~\eqref{eq:boostedq}, is the momentum of the parent boson and therefore the rapidity of the parent boson is just the average of the two hard-parton rapidities $\eta=\left(\eta_1+\eta_2\right)/2$. In these kinematics the transverse momentum is given by
\begin{equation}
    p_{T}^2 = \frac{Q^2}{4}\frac{1-z_{\theta}}{z_{\theta}\sinh^2{\eta}}.
    \label{eq:pT-ztheta-eta}
\end{equation}
Now suppose that the observed hadrons that reach the calorimeters are produced from the hard directions of the parent quarks, carrying momentum fractions $\xi_{1}$ and $\xi_2$, with a small transverse momentum kick
\begin{equation}
    \begin{aligned}
    P_{h_1}^{\mu}&=\xi_1p_1^{\mu}+k_{1\perp}^{\mu}, \\ 
    P_{h_2}^{\mu}&=\xi_2p_2^{\mu}+k_{2\perp}^{\mu}.
    \end{aligned}
    \label{eq:hadron-momenta-factorization}
\end{equation}
Since we are only concerned with the azimuthal component of the transverse momentum kick, let us introduce the transverse basis with respect to the respective hard parton direction $n_i^\mu=p_i^\mu/(p_T\cosh\eta_i)$:
\begin{equation}
\begin{aligned}
k_{i\perp}^\mu&=k_{\eta,i}\,e_{\eta,i}^\mu+k_{\phi,i}\,e_{\phi}^\mu\,,\\
e^\mu_{\eta,i}&\equiv\cosh\eta_i\,\frac{\partial n_i^\mu}{\partial\eta_i}
=\left(0,-\tanh\eta_i\cos\phi_i,\,-\tanh\eta_i\sin\phi_i,\,\mathrm{sech}\,\eta_i\right),\\
e^\mu_{\phi}\footnotemark&\equiv\cosh\eta_1\,\frac{\partial n^\mu_1}{\partial\phi_1}
=\left(0,-\sin\phi_1,\,\cos\phi_1,\,0\right). \label{eq:BasisVectors}
\end{aligned}
\end{equation}
\footnotetext{At leading power the out-of-plane projections of the two partons are antiparallel, $\cosh\eta_2\,\partial n_2^\mu/\partial\phi_2=-e^\mu_\phi$, and therefore we measure both azimuthal recoils with respect to the common normal $e_\phi$.} Note that $e_{\eta,i}^\mu$, which points towards increasing rapidity of parton $i$, parametrizes the recoil within the Born plane, i.e.\ the plane containing the boost axis and the two partons, while $e_\phi^\mu$ is normal to this plane and parametrizes the recoil out of it. At leading power, there will be a small acoplanarity in the detected hadrons' azimuthal angles
\begin{equation}
  \phi_{h_1}-\phi_{h_2}\approx-\pi+\frac{1}{p_{T}}\left(\frac{k_{\phi,1}}{\xi_1}+\frac{k_{\phi,2}}{\xi_2}\right),
  \label{eq:out-of-plane-imbalance}
\end{equation}
which is only sensitive to the out-of-plane recoil. Including soft radiation, the acoplanarity variable defined in Eq.~\eqref{eq:zphi_def} is related to the azimuthal momentum imbalance by
\begin{equation}
    z_\phi \approx \frac{l_\phi^2}{4 p_T^2},  \hspace{2cm} l_\phi^2 = \left(\frac{k_{\phi,1}}{\xi_1}+\frac{k_{\phi,2}}{\xi_2}-\lambda_\phi\right)^2. \label{eq:phi_LP}
\end{equation}
To stay within the leading-power approximation we must require that $k_i/\left(\xi_i p_T\right)\ll1$. For small-recoil, $k_i\sim\Lambda_{\mathrm{QCD}}$, it requires $\xi_ip_T\gg\Lambda_{\mathrm{QCD}}$. This fails for soft hadrons, $\xi_i\lesssim\Lambda_{\mathrm{QCD}}/p_T$, but since the EEC is energy weighted we expect their contribution to be power suppressed. We must nevertheless require $p_T\gg\Lambda_{\mathrm{QCD}}$, which excludes Born configurations aligned with the boost axis. With these kinematics in mind we proceed to show our results for the factorization formula that resums the small-acoplanarity enhancement in the fixed-order calculation. 

As a last side note, in contrast with the fixed-order calculations in Eqs.~\eqref{eq:endpoint_H0}--\eqref{eq:endpoint_H2} we will also keep the hadronic EEC tensors differential in the average azimuth of the detected pair  $\bar{\phi}=\left(\phi_{h_{1}}+\phi_{h_2}\right)/2$ which is related to the azimuthal angle of the quark appearing in Eq.~\eqref{eq:boosted-born-momenta} by 
\begin{equation}
\phi_1 \approx \bar{\phi} - \frac{\pi}{2}
-\frac{1}{2p_{T}}\left(\frac{k_{\phi,1}}{\xi_1}-\frac{k_{\phi,2}}{\xi_2}\right). \label{eq:azimuth_Modulation}
\end{equation}
At leading power in the individual
recoil angles, we evaluate the hard angular factors at
$\phi_1=\bar{\phi}-\pi/2$. This replacement is not implied by small
$z_\phi$ alone, which constrains the difference of the two azimuthal
deflections rather than their average.  The azimuth $\bar{\phi}$ determines the orientation of the detected pair around the boosted boson axis and, for an EEC sourced by a vector boson, will produce further azimuthal modulations.

\subsection{Factorization for $\gamma^*/Z\rightarrow \mathrm{hadrons}$}
The factorized EEC hadronic tensor relevant for $e^{+}e^{-} \rightarrow\gamma^*/Z\rightarrow\mathrm{hadrons}$ and hadronic $Z$ boson decay is
\begin{align}
  \frac{dW^{\mu\nu}_{ij}}{d\zph\,d\zth\,d\bar{\phi}}
  &= N_c\sum_{q}\sum_{h_1,h_2}H \left(\mu,Q\right)\,\frac{Q^4}{32\pi^2}\,
    \frac{|\operatorname{csch}\eta|\,\Theta(z_{\theta}-\sech^2{\eta})}{\zth^{7/2}\sqrt{\zth\cosh^2\eta-1}}
  \int d\xi_1\,d\xi_2\,\xi_1 \xi_2
  \nonumber\\
  &\quad\times \int dl_{\phi}\,\delta\!\left(\zph-\frac{l_\phi^2}{4p_T^2}\right)\int d{k}_{\phi,1}\,dk_{\phi,2}\,d\lambda_{\phi}\;
    \delta\!\left({l}_{\phi}-\frac{{k}_{\phi,1}}{\xi_1}-\frac{{k}_{\phi,2}}{\xi_2}+{\lambda}_{\phi}\right)
  \nonumber\\
  &\quad\times
  \Tr\!\Bigl[\Gamma_{q,i}^\mu\, \Delta _{h_1/q}(\xi_1,k_{\phi,1})\,\Gamma^\nu_{q,j}\, \Delta_{h_2/\bar{q}}(\xi_2,k_{\phi,2})\Bigr]\,
  S_{n_1n_2}({\lambda}_{\phi})\hspace{0.25cm}+ \hspace{0.25cm}\left(n_1\leftrightarrow n_2\right),
  \label{eq:factorization_unsimplified}
\end{align}
with the Dirac structure being $\Gamma^\mu_{f,i}=g_V^{f,i}\,\gamma^\mu+g_A^{f,i}\,\gamma^5\gamma^\mu$ with $ i\in\{\gamma,Z\}$. In this equation $S_{n_1n_2}({\lambda}_{\phi})$ is the EEC soft function~\cite{Moult:2018jzp}, $H(\mu,Q)$ is the hard function~\cite{Moult:2018jzp} and $ \Delta _{h_a/q}(\xi_a,k_{\phi,a}), \hspace{0.2cm} a=1,2$ are the quark and anti-quark TMD Fragmentation Functions (TMDFF)~\cite{Boussarie:2023izj}. Both these distributions are one-dimensional since the factorization formula only requires a small one-dimensional recoil $l_\phi^2 \ll Q^2$, therefore the $k_{\eta,a}$ components are integrated~\cite{Gao:2019ojf,Bhattacharya:2025bqa}:
\begin{align}
    \Delta_{h_a/q}(\xi_a,{k}_{\phi,a})
    &= \frac{1}{4}\int d k_{\eta,a}\,
    \Bigg[
    D_{h_a/q}(\xi_a,k_{a\perp}^2)\gamma_a^+
    - H_{h_a/q}^\perp(\xi_a,k_{a\perp}^2)
    \frac{\sigma_{a,\mu +}k^\mu_{a\perp}}{\xi_a M_{h_a}}
    \Bigg]     \notag\\
    &= \frac{1}{4}\gamma_a^+\int dk_{\eta,a}\,
        D_{h_a/q}(\xi_a,k_{a\perp}^2)
    -    \frac{1}{4} \sigma_{a,\mu +}e^{\mu}_{\phi}\frac{k_{\phi,a}}{\xi_a M_{h_a}}\int d{k}_{\eta,a}\,H_{h_a/q}^\perp(\xi_a,k_{a\perp}^2) \notag\\
    &= \frac{1}{4}\gamma_a^+
        D_{h_a/q}(\xi_a,{k}_{\phi,a})
    -    \frac{1}{4} \sigma_{a,\phi+}\frac{k_{\phi,a}}{\xi_a M_{h_a}}H_{h_a/q}^\perp(\xi_a,{k}_{\phi,a}), \quad a=1,2.
    \label{eq:FF_Dirac_Decomp}
\end{align}
In the equation above the  TMDFF $D_{h_a/q}$ and the Collins function $H^\perp_{h_a/q}$ parametrize the quark TMDFF at leading twist~\cite{Boussarie:2023izj,Boer:2008fr} for an unpolarized final state hadron.
 In the equation above we are using the null directions $n_i = (1,\hat{n}_i)$ and $\bar{n}_a = (1,-\hat{n}_a)$  such that $\gamma_a^+=\gamma_{\mu}n_a^{\mu} $ and $\sigma_{a,\mu+} = \sigma_{\mu \nu}n_a^{\nu}$. The trace in Eq.~\eqref{eq:factorization_unsimplified} is most transparent in a polarization basis aligned with the quark axis (quark helicity frame). To see this, recall that at tree level and in the massless-quark
limit, both vector and axial-vector currents produce a
quark--antiquark pair with opposite helicities. The vector and
axial couplings determine the relative amplitudes of the two
allowed helicity configurations. With spin quantized along the
quark direction, these configurations carry angular momentum
$J_z=\pm1$. Consequently, the EEC hadronic tensor has support only
on the two transverse polarizations in this basis, while the
longitudinal polarization along $e_z^\mu=\tilde n_1^\mu$
decouples:
\begin{equation}
    e_{z,\mu}W^{\mu\nu}=0.
\end{equation}
With this in mind, following~\cite{Boer:2008fr} we introduce a polarization basis to decompose the EEC hadronic tensor:
\begin{equation}
    e_z^\mu =\tilde{n}^\mu_1, \hspace{1cm} e_x^\mu = \frac{B^\mu+ \left(e_z\cdot B\right) e_z^\mu}{\sqrt{1-\left(e_z\cdot B\right)^2}}, \hspace{1cm} e_y^\mu = \frac{1}{Q}\epsilon^{\mu\nu\alpha\beta}q_\nu e_{z,\alpha} e_{x,\beta}, 
\end{equation}
with $B^\mu = \left(\sinh\eta,0,0,\cosh{\eta}\right)$ specified by the boost and $\epsilon^{0123}=1$. The triad above is related to the polarization vectors by $ \epsilon_0^\mu=e_z^\mu $, $\epsilon_{-1}^\mu=-\frac{1}{\sqrt{2}}\left(e_x^\mu+ie_y^\mu\right)$ and $\epsilon_{+1}^\mu=\frac{1}{\sqrt{2}}\left(e_x^\mu-ie_y^\mu\right)$.  
In the kinematics specified by the Born momentum in Eq.~\eqref{eq:boosted-born-momenta} we obtain:
\begin{equation}
\begin{split}
    e_z^\mu = \bigg(&\sinh{\eta} \tanh{\frac{\eta_1-\eta_2}{2}},\,\sech{\frac{\eta_1 - \eta_2}{2}}\cos{\phi_1},\\
    &\sech{\frac{\eta_1 - \eta_2}{2}}\sin{\phi_1},\,\cosh{\eta}\, \tanh{\frac{\eta_1-\eta_2}{2}}\bigg)
\end{split} \label{eq:polarization_z}
\end{equation}\begin{equation}
\begin{split}
    e_x^\mu = \bigg(&\sinh{\eta} \sech{\frac{\eta_1-\eta_2}{2}},\,-\tanh{\frac{\eta_1 - \eta_2}{2}}\cos{\phi_1},\\
    &-\tanh{\frac{\eta_1 - \eta_2}{2}}\sin{\phi_1},\,\cosh{\eta}\, \sech{\frac{\eta_1-\eta_2}{2}}\bigg)
\end{split} \label{eq:polarization_x}
\end{equation}\begin{equation}
    e_y^\mu = e_\phi^\mu. \label{eq:polarization_y}
\end{equation}
The trace in Eq.~\eqref{eq:factorization_unsimplified} is given by
{\allowdisplaybreaks
\begin{align}
&\Tr\!\Bigl[
  \Gamma_i^\mu \Delta_{h_1/q}(\xi_1,{k}_{\phi,1})\,
  \Gamma_j^\nu \Delta_{h_2/\bar q}(\xi_2,{k}_{\phi,2})
+ n_1\leftrightarrow n_2 \, \Bigr]
\notag\\*
&=\frac{z_\theta}{2}\Biggl[ \left(g_V^ig_V^j+g_A^ig_A^j\right)
  \left(D_{h_1/q}(\xi_1,k_{\phi,1})
  D_{h_2/\bar{q}}(\xi_2,k_{\phi,2}) + q\leftrightarrow\bar{q}\,\right)
  \bigl(e_x^\mu e_x^\nu+e_y^\mu e_y^\nu\bigr)
\notag\\
&\quad+
 \left(g^i_V g^j_V-g_A^i g_A^j\right)\frac{
    {k}_{\phi,1}{k}_{\phi,2}
  }{
    \xi_1\xi_2M_{h_1}M_{h_2}
  }\left(H_{h_1/q}^\perp(\xi_1,k_{\phi,1})
    H_{h_2/\bar{q}}^\perp(\xi_2,k_{\phi,2})+ q\leftrightarrow \bar{q}\,\right)
\notag\\
&\qquad\times
  \bigl(e_x^\mu e_x^\nu-e_y^\mu e_y^\nu\bigr)
\notag\\
&\quad-i\left(g^i_Vg^j_A + g^i_Ag^j_V\right)  \left( D_{h_1/q}(\xi_1,k_{\phi,1})
D_{h_2/\bar{q}}(\xi_2,k_{\phi,2})-q\leftrightarrow \bar{q} \, \right)
  \bigl(e_x^\mu e_y^\nu-e_y^\mu e_x^\nu\bigr)
  \notag\\
&\quad
-i \left(g^i_Vg^j_A - g^i_Ag^j_V\right) \frac{
    {k}_{\phi,1}{k}_{\phi,2}
  }{
    \xi_1\xi_2M_{h_1}M_{h_2}
  }\left(H_{h_1/q}^\perp(\xi_1,k_{\phi,1})
    H_{h_2/\bar{q}}^\perp(\xi_2,k_{\phi,2})- q\leftrightarrow \bar{q}\,\right)
\notag\\
&\qquad\times
  \bigl(e_x^\mu e_y^\nu+e_y^\mu e_x^\nu\bigr)
\Biggr]. \label{eq:Trace}
\end{align}
}
Charge-conjugation invariance of QCD implies $D_{h/q}=D_{\bar h/\bar q}$ and $H^\perp_{h/q}=H^\perp_{\bar h/\bar q}$. Consequently, the imaginary terms cancel in the fully inclusive
hadron sum of Section~\ref{sec:InclusiveEEC}, as well as in the
charge-conjugation-symmetric like-sign and unlike-sign pion-pair
sums of Section~\ref{sec:Zdecaypheno}. We therefore omit them below.
Therefore we decompose Eq.~\eqref{eq:factorization_unsimplified} as
\begin{equation}
    \frac{dW^{\mu\nu}_{ij}}{d\zph\,d\zth\,d\bar{\phi}} =
    \left(\epsilon_{+1}^\mu\epsilon_{+1}^{\nu*}+\epsilon_{-1}^\mu\epsilon_{-1}^{\nu*}\right)W^{\mathrm{U}}_{ij}
    +\left(\epsilon_{+1}^\mu\epsilon_{-1}^{\nu*}+\epsilon_{-1}^\mu\epsilon_{+1}^{\nu*}\right)W^{\perp}_{ij}, \label{eq:ZEEC_polarization_decomp}
\end{equation}where the superscripts $\mathrm{U}$ and $\perp$ anticipate that the two coefficients are built from the unpolarized and Collins-type jet functions, $J$ and $J^\perp$, defined below. The first term projects into the helicity diagonal part of the $\gamma^*/Z$ density matrix and is insensitive to the quark polarization. The second term is sensitive to the superposition of quark-antiquark states with opposite helicities, $q_{+} \bar{q}_{-}\rightarrow  J_z = +1$ or $q_{-} \bar{q}_{+} \rightarrow J_z = -1$,  that are produced by the current. This coherence generates the
transverse-spin correlations between the quark and antiquark typical of the Collins effect. 

We write the TMDFFs in position space with the conventions
\begin{equation}
    D_{h/i}(\xi,\tilde{b}_{\perp}) = \int \frac{d\tilde{k}_{\perp }}{\xi} e^{-i \tilde{k}_{\perp}\tilde{b}_\perp/\xi}  D_{h/i}(\xi,\tilde{k}_{\perp}) ,
\end{equation}
\begin{equation}
    i \frac{\tilde{b}_\perp}{2}\tilde{H}^\perp_{h/i}(\xi,\tilde{b}_{\perp}) = \int \frac{d\tilde{k}_{\perp }}{\xi} e^{-i \tilde{k}_{\perp}\tilde{b}_\perp/\xi}   \frac{\tilde{k}_{\perp}}{\xi M_{h}}H^{\perp}_{h/i}(\xi,\tilde{k}_{\perp})\label{eq:CollinsFT},
\end{equation}
and define the EEC jet function 
\begin{equation}
    J_{i}\left(b,\mu,\zeta_i/\nu^2\right)\equiv \sum_{h}\int d\xi\, \xi^2 D_{h/i}\left(\xi,b,\mu,\zeta_i/\nu^2\right), 
\end{equation}
and the Collins-type jet function
\begin{equation}
    J_i^\perp\left(b,\mu,\zeta_i/\nu^2\right) \equiv \sum_{h}\int d\xi\,\xi^2 \tilde{H}^\perp_{h/i}\left(\xi,b,\mu,\zeta_i/\nu^2\right).
\end{equation}
The relevant coupling constants are given by
\begin{equation}
    g_V^{f,\gamma}=e\,Q_f,\qquad g_A^{f,\gamma}=0,\qquad
    g_V^{f,Z}=\frac{e}{2s_Wc_W}\,v_f,\qquad g_A^{f,Z}=\frac{e}{2s_Wc_W}\,a_f,
\end{equation}
\begin{equation}
    v_f=T_3^f-2Q_f\,s_W^2,\qquad a_f=T_3^f,
\end{equation}
where $s_W=\sin\theta_W$, $c_W=\cos\theta_W$, $Q_f$ is the electric charge in units of $e$ and $T_3^f$ the third component of weak isospin. The propagators are
\begin{equation}
    P_\gamma(Q^2)=\frac{1}{Q^2},\qquad
    P_Z(Q^2)=\frac{1}{Q^2-M_Z^2+iM_Z\Gamma_Z}. \label{eq:Propagators}
\end{equation}
Finally, we define
\begin{equation}
    c^{f,\pm}_{ij}=g_V^{f,i}g_V^{f,j}\pm g_A^{f,i}g_A^{f,j}.
\end{equation}
With these definitions the two coefficients in Eq.~\eqref{eq:ZEEC_polarization_decomp} are
\begin{equation}
\begin{split}
W^{\mathrm{U}}_{ij}&=N_c\sum_q c^{q,+}_{ij}\,H(\mu,Q)\,\frac{Q^4p_T}{16\pi^3}\,
\frac{\mathcal{K}(z_\theta;\eta) }{\sqrt{z_\phi}} \\
&\quad\times\int_0^\infty db\,\cos\!\big(2bp_T\sqrt{z_\phi}\big)\,S_{n_1n_2}(b,\mu,\nu)\,
J_q\big(b,\mu,\zeta_1/\nu^2\big)\,J_{\bar{q}}\big(b,\mu,\zeta_2/\nu^2\big), \label{eq:HUnpol_ij}
\end{split}
\end{equation}
\begin{equation}
\begin{split}
W^{\perp}_{ij}&=N_c\sum_q c^{q,-}_{ij}\,H(\mu,Q)\,\frac{Q^4p_T}{16\pi^3}\,
\frac{\mathcal{K}(z_\theta;\eta) }{\sqrt{z_\phi}} \\
&\quad\times\frac{1}{4}\int_0^\infty db\,b^2\cos\!\big(2bp_T\sqrt{z_\phi}\big)\,S_{n_1n_2}(b,\mu,\nu)\,
J^\perp_q\big(b,\mu,\zeta_1/\nu^2\big)\,J^\perp_{\bar{q}}\big(b,\mu,\zeta_2/\nu^2\big).\label{eq:HCollins_ij}
\end{split}
\end{equation}
The EEC for hadrons produced in $e^+e^-$ collisions (Eq.~\ref{eq:epem_eec}) can be easily obtained by contracting Eq.~\eqref{eq:ZEEC_polarization_decomp} with the spin-averaged leptonic tensor and summing over the couplings to the vector bosons. Since the decomposition we obtained of the EEC hadronic tensor is symmetric under $\mu \leftrightarrow \nu$, we only have to worry about contracting it with the symmetric part of the spin-averaged leptonic tensor $L_{\mu\nu}^{ij} = -c_{ij}^{e,+}\,Q^2/2\left(g_{\perp\mu\nu}+\hat{l}_\mu\hat{l}_\nu\right)$ where $\hat{l}^\mu = \left(l_1^\mu-l_2^\mu\right)/Q$, $l_{12}$ are the momenta of the incoming beams and $ c^{e,+}_{ij}=g_V^{e,i}g_V^{e,j}+ g_A^{e,i}g_A^{e,j}$ gives the coupling of the leptons to the virtual photon and Z-boson including their possible interference. Before showing the final results in terms of $z_\phi,z_\theta$ and $\bar{\phi}$, it is useful to carry out the contraction of Eq.~\eqref{eq:ZEEC_polarization_decomp} with the leptonic tensor in the vector boson center-of-mass frame:
\begin{equation}
\begin{split}
L_{\mu\nu}^{ij}\,
\frac{dW^{\mu\nu}_{ij}}{dz_\phi\,dz_\theta\,d\bar{\phi}}
&= c^{e,+}_{ij}\frac{Q^2}{2}
\Bigl\{
  \bigl[1+(\hat{l}\cdot\vec{e}_z)^2\bigr]W^{\mathrm{U}}_{ij}
  +\bigl[(\hat{l}\cdot\vec{e}_x)^2-(\hat{l}\cdot\vec{e}_y)^2\bigr]
   W^{\perp}_{ij}
\Bigr\}
\\
&= c^{e,+}_{ij}\frac{Q^2}{2}
\Bigl\{
  (1+\cos^2\Theta)W^{\mathrm{U}}_{ij}
  +\sin^2\Theta\cos(2\chi)\,W^{\perp}_{ij}
\Bigr\}.
\end{split}
\end{equation}
Here $\vec{e}_x$, $\vec{e}_y$ and $\vec{e}_z$ are the spatial parts of the triad we defined in Eqs.~\eqref{eq:polarization_z}--\eqref{eq:polarization_y}  evaluated in the vector boson center-of-mass frame. $\Theta$ is the angle between $\hat{l}$ and $\vec{e}_z$, and $\chi$ is the azimuth of $\hat{l}$ about $\vec{e}_z$, measured from $\vec{e}_x$ toward $\vec{e}_y$. These modulations match the familiar Collins angular
structure~\cite{Boer:2008fr,Kang:2023big,Boer:1997mf} in the Gottfried--Jackson (GJ) frame~\cite{Gottfried:1964nx,Pitonyak:2013dsu}, where the vector boson is at rest and the $z$-axis points along the momentum of
one detected hadron. After this sanity check, we rewrite this expression now in terms of our boosted variables and normalize. We find
\begin{equation}
\begin{split}
\frac{d\Sigma_{e^+e^-}}{d\zph\,d\zth\,d\bar{\phi}}
&= \frac{\sech^2\eta}{8Q^2\sigma_0}\sum_{i,j\in\{\gamma,Z\}}{\rm Re}\big[P_iP_j^*\big]\,c^{e,+}_{ij}\bigg\{\\
&\qquad
\Big[\big(1+\cos^2\theta^*\big)\big(1+\cos^2\theta_l\big)+2\sin^2\theta^*\sin^2\theta_l\Big]W^{\mathrm{U}}_{ij}\\
&\qquad
+\sin^2\theta^*\big(3\cos^2\theta_l-1\big)\,W^{\perp}_{ij}\\
&\qquad
-\sin^2\theta_l\cos 2\big(\bar{\phi}-\phi_l\big)
\Big[\sin^2\theta^*\,W^{\mathrm{U}}_{ij}+\big(1+\cos^2\theta^*\big)W^{\perp}_{ij}\Big]\bigg\}, \label{eq:epem_EEC}
\end{split}
\end{equation}where $\theta_l$ is the polar angle of the beam in the virtual-boson center-of-mass frame, $\phi_l$ is its azimuthal angle and $\theta^*$ is the polar angle of one of the hard quarks in the same frame. In terms of the boosted variables we have
\begin{equation}
    \sin^2{\theta^*}= \frac{1-z_\theta}{z_\theta \sinh^2{\eta}},\hspace{2cm} \cos^2\theta^* = \frac{z_\theta \cosh^2\eta -1}{z_\theta \sinh^2\eta}. \label{eq:sinCOM}
\end{equation}
Azimuthal $\cos{2\left(\bar{\phi}-\phi_l\right)}$ modulations, which depend on the orientation of the correlated pair in the plane transverse to the boost axis, are present in both the unpolarized and Collins terms. For small $\sin^2\theta^*\ll1$ (which corresponds to the $z$-axis being aligned with the quark momentum in the virtual-boson center-of-mass frame) the azimuthal modulation of the unpolarized temrs is suppressed. However, at fixed $Q^2$ taking arbitrarily small $\sin^2\theta^*$ this would imply small $p_T/Q\ll1$ which violates the requirement of $p_T\gg \Lambda_{\mathrm{QCD}}$ that justifies the leading-power factorization formula, as discussed around Eq.~\eqref{eq:phi_LP}.  Furthermore, in head-on $e^+e^-$ collisions, as at PEP-II, $\theta_l =0$ up to initial-state radiation, so all the azimuthal modulations vanish. Even then, sensitivity to the Collins effect remains non-zero through the second term in Eq.~\eqref{eq:epem_EEC}.

The particular form of the azimuthal modulations in Eq.~\eqref{eq:epem_EEC} was determined by the contraction of the leptonic tensor with the EEC hadronic tensor. For an EEC computed from the hadronic decay of a boosted boson (for recent work in this direction, see~\cite{Gao:2026xuq,Gao:2026hoj,Ricci:2022htc,Holguin:2026vld}) the EEC hadronic tensor would instead be contracted with the density matrix of the vector boson. In this setting, which might be more natural for this boosted formalism, one can study whether the resulting azimuthal modulations can help isolate the Collins effect. We test this idea in Section~\ref{sec:Zdecaypheno}. Here we prepare the formalism by obtaining the factorization formula for the hadronic decay of a $Z$ boson under the narrow-width approximation in which production and decay of the vector boson factorize. The decay rate is given in terms of a trace over the $Z$-boson polarizations. The EEC therefore follows from Eq.~\eqref{eq:ZEEC_polarization_decomp} with $i=j=Z$ and $Q=M_Z$, by
contracting with the $Z$-boson density matrix. Let us write the hadronic EEC for $Z$ decay in the polarization basis
\begin{equation}
    (W_{ZZ} )_{\lambda\lambda'}=
    \begin{pmatrix}
        W_{ZZ}^{\mathrm{U}} & 0 & W_{ZZ}^{\perp}\\
        0 & 0 & 0\\
        W_{ZZ}^{\perp} & 0 & W_{ZZ}^{\mathrm{U}}
    \end{pmatrix}_{\lambda,\lambda'=(+,0,-)},
\end{equation}
where $W^{\mathrm{U}}_{ZZ}$ and $W^{\perp}_{ZZ}$ are given in Eqs.~\eqref{eq:HUnpol_ij} and
\eqref{eq:HCollins_ij}. Then the hadronic $Z$-decay EEC is given by
\begin{equation}
\begin{split}
    \frac{d\Sigma_Z}{d\zph\,d\zth\,d\bar\phi}
    &=\frac{\sech^2\eta}{2M_Z^3\Gamma_{Z,0}}\,\Tr\big[\rho' W_{ZZ}\big]\\
    &=\frac{\sech^2\eta}{2M_Z^3\Gamma_{Z,0}}\Big[\big(\rho'_{++}+\rho'_{--}\big)W_{ZZ}^{\mathrm{U}}
    +\big(\rho'_{+-}+\rho'_{-+}\big)W_{ZZ}^{\perp}\Big],
\end{split} \label{eq:Z_fact}
\end{equation}
with $\Gamma_{Z,0}$ being the $Z$ boson hadronic decay width at LO. Here $\rho'$ is the density matrix of the produced $Z$ boson, normalized to $\Tr\rho'=1$ and quantized
along $\tilde n_1$, i.e. in the quark helicity frame built from Eqs.~\eqref{eq:polarization_z}--\eqref{eq:polarization_y}. If the density
matrix $\rho$ is given in a different frame, for example the vector boson helicity frame, then
$\rho'=D^{(1)\dagger}(\alpha,\beta,\gamma)\,\rho\,D^{(1)}(\alpha,\beta,\gamma)$, where $D^{(1)}(\alpha,\beta,\gamma)$ is the spin-1 Wigner matrix connecting the polarization vectors in the two different frames. To illustrate this observable in Section~\ref{sec:Zdecaypheno} we study photon-tagged $Z$ hadronic decay in $e^+e^-\rightarrow Z\gamma$ collisions by constructing the necessary density matrix and writing the full factorization formula in terms of it.

\subsection{Factorization for $W^{\pm}\rightarrow \mathrm{hadrons}$}
As reviewed in Section~\ref{sec:Decompositions}, in the massless quark limit the hadronic EEC with currents coupling a $W^+$ to quarks (Eq.~\eqref{eq:Ws_currents}) is analogous to Eq.~\eqref{eq:factorization_unsimplified}, with the modification that the inserted currents only couple to left-handed quarks (and right-handed anti-quarks) of different flavors. A consequence of this is that the produced hard-pair is not spin-entangled; there is only one allowed helicity configuration which has both particles longitudinally polarized. It follows that there is no transverse spin correlation to couple to the Collins TMDFF~\cite{Arratia:2022oxd,Gao:2026hoj}. In other words, the would-be-Collins effect for this particular EEC hadronic tensor is proportional to $(g_V^W)^2-(g_A^W)^2= 0 $ (this can be read directly from Eq.~\eqref{eq:Trace} by taking $q\leftrightarrow u$ and $\bar{q}\leftrightarrow \bar{d}$ and plugging in the W-couplings to these quarks into the $i,j$ indices). The resulting  hadronic $W^{\pm}$-decay EEC is given by
\begin{equation}
    \frac{d\Sigma_{W^\pm}}{d\zph\,d\zth\,d\bar{\phi}}
    =\frac{\sech^2\eta}{2M_W^3\Gamma_{W,0}}\,\big(\rho'_{++}+\rho'_{--}\big)\,W^{\mathrm{U}}_{++},
\end{equation}
where $W^{\mathrm{U}}_{++}$ is given by Eq.~\eqref{eq:HUnpol_ij} with $c^{+}_{WW}=g^2|V_{ud}|^2/4$, the flavor sum
running over $u_i\bar d_j$ and with the jet functions $J_{u_i}J_{\bar d_j}$.
\subsection{Factorization for $H\rightarrow \mathrm{hadrons}$}

We now consider the EEC measured from the hadronic decay of a Higgs boson through the quark and gluon channels. Let us begin with the quark channel. The Yukawa coupling produces a quark antiquark pair in the triplet $m=0$ state which allows for transverse correlations. Therefore we will find that even though the Higgs itself is a scalar particle, measuring its hadronic decay products still gives sensitivity to the Collins effect. The factorization formula reads
\begin{align}
  \frac{dW_{YY}}{d\zph\,d\zth\,d\bar{\phi}}
  &= N_c\sum_{q}\frac{y_q^2}{2}\sum_{h_1,h_2}H^{Y} \left(\mu,M_H\right)\,\frac{M_H^4}{32\pi^2}\,
    \frac{|\operatorname{csch}\eta|\,\Theta(z_{\theta}-\sech^2{\eta})}{\zth^{7/2}\sqrt{\zth\cosh^2\eta-1}}
  \int d\xi_1\,d\xi_2\,\xi_1 \xi_2
  \nonumber\\
  &\quad\times \int dl_{\phi}\,\delta\!\left(\zph-\frac{l_\phi^2}{4p_T^2}\right)\int d{k}_{\phi,1}\,dk_{\phi,2}\,d\lambda_{\phi}\;
    \delta\!\left({l}_{\phi}-\frac{{k}_{\phi,1}}{\xi_1}-\frac{{k}_{\phi,2}}{\xi_2}+{\lambda}_{\phi}\right)
  \nonumber\\
  &\quad\times
  \Tr\!\Bigl[\Delta _{h_1/q}(\xi_1,k_{\phi,1}) \Delta_{h_2/\bar{q}}(\xi_2,k_{\phi,2})\Bigr]\,
  S_{n_1n_2}({\lambda}_{\phi})\hspace{0.25cm}+ \hspace{0.25cm}\left(n_1\leftrightarrow n_2\right),
  \label{eq:factorizatiHiggsqq}
\end{align}Here $H^Y(\mu,M_H)$ is the scalar (Yukawa) hard function~\cite{Gao:2019mlt}. The scalar nature of the currents yields a simpler trace
\begin{align}
&\Tr\!\Bigl[\Delta_{h_1/q}(\xi_1,{k}_{\phi,1})\,
  \Delta_{h_2/\bar q}(\xi_2,{k}_{\phi,2})
+ n_1\leftrightarrow n_2 \, \Bigr]
\notag\\
&=\frac{z_\theta}{2}\Biggl[
  \left(D_{h_1/q}(\xi_1,k_{\phi,1})
  D_{h_2/\bar{q}}(\xi_2,k_{\phi,2}) + q\leftrightarrow\bar{q}\,\right)
\notag\\
&\quad-
 \frac{
    {k}_{\phi,1}{k}_{\phi,2}
  }{
    \xi_1\xi_2M_{h_1}M_{h_2}
  }\left(H_{h_1/q}^\perp(\xi_1,k_{\phi,1})
    H_{h_2/\bar{q}}^\perp(\xi_2,k_{\phi,2})+ q\leftrightarrow \bar{q}\,\right)\, 
\Biggr]. \label{eq:TraceHqq}
\end{align}
Indeed since there are no polarization vectors to worry about, we can immediately write $H\to q\bar q$ EEC as
\begin{equation}
    \frac{d\Sigma_{H\to q\bar{q}}}{d\zph\,d\zth}
    =\frac{\pi\sech^2\eta}{2M_H^3\Gamma^{q\bar{q}}_{H,0}}\left[ W^{\mathrm{U}}_{YY}+ W^{\perp}_{YY} \right],
\end{equation}
where $W_{YY}^{\mathrm{U}}$ is given by replacing $H(\mu,Q)\rightarrow H^{Y}(\mu,M_H)$ and $c_{ij}^{q,+}\leftrightarrow  y_q^2$ in Eq.~\eqref{eq:HUnpol_ij} and $W_{YY}^{\perp}$ is given by doing the same replacements in Eq.~\eqref{eq:HCollins_ij} and $\Gamma^{q\bar{q}}_{H,0}$ is the LO $H\rightarrow q\bar{q}$ width.

The gluonic decay channel involves gluon TMD fragmentation functions instead, and the factorization formula now reads
\begin{align}
  \frac{dW_{GG}}{d\zph\,d\zth\,d\bar{\phi}}
  &= \left(N_c^2-1\right)C_g^2\sum_{h_1,h_2}H^{G}\left(\mu,Q\right)\,\frac{M_H^6}{16\pi^2}\,
    \frac{|\operatorname{csch}\eta|\,\Theta(z_{\theta}-\sech^2{\eta})}{\zth^{5/2}\sqrt{\zth\cosh^2\eta-1}}
  \int d\xi_1\,d\xi_2\,\xi_1 \xi_2
  \nonumber\\
  &\quad\times \int dl_{\phi}\,\delta\!\left(\zph-\frac{l_\phi^2}{4p_T^2}\right)\int d{k}_{\phi,1}\,dk_{\phi,2}\,d\lambda_{\phi}\;
    \delta\!\left({l}_{\phi}-\frac{{k}_{\phi,1}}{\xi_1}-\frac{{k}_{\phi,2}}{\xi_2}+{\lambda}_{\phi}\right)
  \nonumber\\
  &\quad\times
  \hat{V}_{\mu\nu} \Delta_{h_1/g}^{\mu \mu'}\left(\xi_1,k_{\phi,1}\right)\left(\hat{V}_{\mu'\nu'}\right)^*\Delta^{\nu \nu'}_{h_2/g}\left(\xi_2,k_{\phi,2}\right)S_{n_1n_2}^A\left(\lambda_\phi\right) \;+\;(n_1\leftrightarrow n_2)\,,
  \label{eq:factorization_Hgg}
\end{align}
where $C_g$ is defined through $\mathcal{L}_{\rm eff}=-C_g\,h\,G^b_{\mu\nu}G^{b\,\mu\nu}$, with $C_g=-\alpha_s/(12\pi v)$ at leading order in the heavy-top limit (where $v$ is the electroweak vacuum expectation value) and $\hat{V}^{\mu\nu} = g_{T,12}^{\mu\nu}+p_1^\mu p_2^\nu/\left(p_1\cdot p_2\right)$ is the normalized projector coming from this effective vertex with $g_{T,12}^{\mu\nu} = \eta^{\mu\nu}-\left(p_1^\mu p_2^\nu+p_1^\nu p_2^\mu\right)/\left(p_1\cdot p_2\right)$. $S_{n_1n_2}^A\left(\lambda_\phi\right)$ is the EEC adjoint soft function~\cite{Ebert:2020sfi} and $\Delta_{h_i/g}$ are the gluon TMDFF~\cite{Boussarie:2023izj} which are one-dimensional since the recoil is purely azimuthal
\begin{equation}
\begin{split}
    \Delta_{h_i/g}^{\mu\nu}\left(\xi_i,k_{\phi,i}\right) &= \frac{1}{2}\int dk_{\eta,i} \left[-g_{iT}^{\mu\nu}D_{h_i/g}\left(\xi_i,k_{\perp i}\right)+\frac{k_{\perp i}^2}{\xi_i^2 M_{h_i}^2}\left(\frac{g_{iT}^{\mu\nu}}{2}+\frac{k_{\perp i}^\mu k_{\perp i}^\nu}{k_{\perp i}^2}\right)H^{\perp}_{h_i/g}\left(\xi_i,k_{\perp i}\right)\right]\\
    &= -\frac{1}{2}\,g_{iT}^{\mu\nu}\,D_{h_i/g}\left(\xi_i,k_{\phi,i}\right)+\frac{1}{4}\left(e_\phi^\mu e_\phi^\nu-e_{\eta,i}^\mu e_{\eta,i}^\nu\right)\bar H^{\perp}_{h_i/g}\left(\xi_i,k_{\phi,i}\right)\\
    &= \frac{1}{2}\left(e_{\eta,i}^\mu e_{\eta,i}^\nu+e_\phi^\mu e_\phi^\nu\right)D_{h_i/g}\left(\xi_i,k_{\phi,i}\right)+\frac{1}{4}\left(e_\phi^\mu e_\phi^\nu-e_{\eta,i}^\mu e_{\eta,i}^\nu\right)\bar H^{\perp}_{h_i/g}\left(\xi_i,k_{\phi,i}\right).
\end{split}
\end{equation}
The first term is proportional to the unpolarized gluon TMDFF and the second to the linearly polarized gluon TMDFF. In the last line we wrote transverse metric in terms of the unit vectors in Eq.~\eqref{eq:BasisVectors}  as $g_{iT}^{\mu\nu}=-\left(e_{\eta,i}^\mu e_{\eta,i}^\nu+e_\phi^\mu e_\phi^\nu\right)$ and we defined the one-dimensional TMDFFs by
\begin{align}
    D_{h_i/g}\left(\xi_i,k_{\phi,i}\right)&\equiv\int dk_{\eta,i}\,D_{h_i/g}\left(\xi_i,k_{\perp i}\right),\notag\\
    \bar H^{\perp}_{h_i/g}\left(\xi_i,k_{\phi,i}\right)&\equiv\int dk_{\eta,i}\,\frac{k_{\phi,i}^2-k_{\eta,i}^2}{\xi_i^2M_{h_i}^2}\,H^{\perp}_{h_i/g}\left(\xi_i,k_{\perp i}\right).
    \label{eq:OnedimGluonFF}
\end{align}
To simplify Eq.~\eqref{eq:factorization_Hgg} it is easiest to notice that $\hat{V}^{\mu\nu}$ and $g_{iT}^{\mu\nu}$ all reduce to the same tensor $g_{T,12}^{\mu\nu}$ up to terms proportional to $n_i^\mu$ which vanish when contracted since at leading order the gluon TMDFFs are transverse, $n_{i\mu}\Delta^{\mu\nu}_{h_i/g}=0$. It follows that 
\begin{equation}
\begin{split}
&\hat V_{\mu\nu}\,\Delta^{\mu\mu'}_{h_1/g}\left(\xi_1,k_{\phi,1}\right)\big(\hat V_{\mu'\nu'}\big)^*\,\Delta^{\nu\nu'}_{h_2/g}\left(\xi_2,k_{\phi,2}\right)\;+\;(n_1\leftrightarrow n_2)\\
&\qquad=\,D_{h_1/g}\left(\xi_1,k_{\phi,1}\right)D_{h_2/g}\left(\xi_2,k_{\phi,2}\right)
+\frac{1}{4} \bar{H}^{\perp}_{h_1/g}\left(\xi_1,k_{\phi,1}\right) \bar{H}^{\perp}_{h_2/g}\left(\xi_2,k_{\phi,2}\right).
\end{split}
\end{equation}
We express the factorization formula in position space with the conventions
\begin{align}
D_{h_i/g}\left(\xi_i,b_\phi\right)&=\int\frac{dk_{\phi,i}}{\xi_i}\,e^{-ik_{\phi,i} b_\phi/\xi_i}\,D_{h_i/g}\left(\xi_i,k_{\phi,i}\right),\\[4pt]
\left(\frac{i b_\phi}{2}\right)^{2}\tilde H^{\perp}_{h_i/g}\left(\xi_i, b_\phi\right)&=\int\frac{dk_{\phi,i}}{\xi_i}\,e^{-ik_{\phi,i} b_\phi/\xi_i}\,\bar H^{\perp}_{h_i/g}\left(\xi_i,k_{\phi,i}\right),\label{eq:Pos_LingPol}
\end{align}
and define the gluon EEC jet function and the linearly polarized gluon EEC jet function
\begin{align}
J_g\left(b,\mu,\zeta/\nu^2\right)&\equiv\sum_{h}\int d\xi\,\xi^2\,D_{h/g}\left(\xi,b,\mu,\zeta/\nu^2\right),\notag\\
J^{\perp}_g\left(b,\mu,\zeta/\nu^2\right)&\equiv\sum_{h}\int d\xi\,\xi^2\,\tilde H^{\perp}_{h/g}\left(\xi,b,\mu,\zeta/\nu^2\right).
\end{align}
The resulting factorization formulas read
\begin{align}
W^{\mathrm{U}}_{GG}
=\;& C_g^2(N_c^2-1)\,H^{G}(\mu,Q)\,\frac{M_H^6\,p_T}{8\pi^3}\,\frac{\mathcal{K}(z_\theta;\eta)}{\sqrt{z_\phi}}
\int_0^\infty db\,\cos\!\big(2bp_T\sqrt{z_\phi}\big)\,S^A_{n_1n_2}(b,\mu,\nu)\nonumber\\
&\times J_g\big(b,\mu,\zeta_1/\nu^2\big)\,J_g\big(b,\mu,\zeta_2/\nu^2\big)\,,\label{eq:HggU}\\[4pt]
W^{\perp}_{GG}
=\;&C_g^2(N_c^2-1)\,H^{G}(\mu,Q)\,\frac{M_H^6\,p_T}{8\pi^3}\,\frac{\mathcal{K}(z_\theta;\eta)}{\sqrt{z_\phi}}
\int_0^\infty db\,\frac{b^4}{64}\,\cos\!\big(2bp_T\sqrt{z_\phi}\big)\,S^A_{n_1n_2}(b,\mu,\nu)\nonumber\\
&\times J^\perp_g\big(b,\mu,\zeta_1/\nu^2\big)\,J^\perp_g\big(b,\mu,\zeta_2/\nu^2\big)\,.\label{eq:HggL}
\end{align}
From which the hadronic gluon channel Higgs-decay EEC is given by
\begin{equation}
\frac{d\Sigma_{H\to gg}}{dz_\phi dz_\theta}
=\frac{\pi\,\operatorname{sech}^2\eta}{M_H^3\,\Gamma^{gg}_{H,0}}\Big[W^{\mathrm{U}}_{GG}+W^{\perp}_{GG}\Big]\,,
\label{eq:Hggfactfinal}
\end{equation}
where $\Gamma^{gg}_{H,0}$ is the LO $H\to gg$ width.

\subsection{Discussion of results at $\eta=0$}

Let us take a moment to discuss how to take the $\eta=0$ limit in the fixed-order results in Eqs.~\eqref{eq:endpoint_H0}--\eqref{eq:HiggsGluonFO}  and in the factorization formulas we obtained. First note that even though the transverse momentum of the hard partons is invariant under the boost that takes them back to the COM frame, naively plugging in $\eta=0$ and $z_\theta=1$ into Eq.~\eqref{eq:pT-ztheta-eta} is not a well-defined operation. This is not surprising since the kinematic condition to justify our expansion, Eq.~\eqref{eq:kinematic_region}, is valid for non-zero boson rapidity. We can still reach the COM frame by first making our observables differential in $\sin^2\theta^*$ (defined in Eq.~\eqref{eq:sinCOM}) such that the parton transverse momentum is explicitly boost invariant: $p_T^2(\sin^2\theta^*)=Q^2/4 \sin^2{\theta^*}$. For instance with these variables we could now easily take $\eta=0$ in Eq.~\eqref{eq:endpoint_H0}
\begin{align}
\frac{dW_0(\sin^2\theta^*,z_\phi;\eta)}{d\sin^2\theta^*\,dz_\phi}={}&\frac{N_cQ^4}{8\pi}\,\frac{1+\sinh^2\eta\sin^2\theta^*}{|\cos\theta^*|}\nonumber\\
&\times\left\{\delta(z_\phi)-\frac{\alpha_sC_F}{2\pi}\,\frac{1}{z_\phi}\left[3+2\log\!\left(4z_\phi\sin^2\theta^*\right)\right]\right\}+\mathcal O\!\left(\alpha_sz_\phi^{-1/2}\right). \label{eq:fixed_order_COM}
\end{align}
This is a natural change of variables. In the COM frame the boson carries no spatial momentum, therefore there is no preferred direction with respect to which to define the $z$-axis; we may choose it arbitrarily, for instance along the beam or along any fixed reference direction. Once this choice is made, the information on the orientation of the back-to-back pair relative to the axis is carried by $\theta^*$, the polar angle of the hard partons, while $z_\theta$ degenerates: at the Born level every orientation is mapped to $z_\theta=1$. For $\eta\neq0$ the boost axis is physical and the aberration encodes $\theta^*$ in the opening angle through Eq.~\eqref{eq:sinCOM}, which is why $z_\theta$ is a good variable in the boosted frame, but not in the limit $\eta\to0$. Since both $p_T$ and $z_\phi$ are defined transverse to the same axis, and are therefore invariant under boosts along it, $\sin^2\theta^*$ is the variable that interpolates smoothly between the boosted and the COM frames for our two dimensional EEC.

One might still be suspicious of this change of variables since Eq.~\eqref{eq:endpoint_H0} is only justified under the assumptions of the inequality in~\eqref{eq:kinematic_region}. This result can be confirmed independently by working directly in the COM frame, with the distribution differential in $\sin^2\theta^*$ and $z_\phi$ given by
\begin{align}
\frac{dW_0(\sin^2\theta^*,z_\phi;\eta=0)}{d\sin^2\theta^*\,dz_\phi}={}&\frac{1}{2}\int d\Omega_1\,d\Omega_2\,\frac{dW_0(z)}{d\Omega_1\,d\Omega_2}\,\delta\!\left(z_\phi-\frac{1+\cos(\phi_1-\phi_2)}{2}\right)\nonumber\\
&\times\left[\delta\!\left(\sin^2\theta^*-\sin^2\theta_1\right)+\delta\!\left(\sin^2\theta^*-\sin^2\theta_2\right)\right]. \label{eq:DefCOM}
\end{align}
The factor $1/2$ together with the sum of the two delta functions symmetrizes over the two particles in the correlated pair. In this case the leading contribution as $z_\phi \ll1$ comes from the region $4z_\phi\sin^4\theta^*\sim x^2 \ll 1$ with $x\equiv \cos{\theta_2}+\cos{\theta^*} \rightarrow0$. Expanding $z=(1-\hat{n}_1\cdot\hat{n}_2)/2$ in this limit gives
\begin{equation}
    \frac{1}{1-z} \approx \frac{4\sin^2{\theta^*}}{4z_\phi\sin^4{\theta^*}+x^2}.
\end{equation}
This integration region is expected since it follows from $z_\phi\sim \left(\tilde{\eta}-\eta\right)^2\ll1$ at $\eta=0$ by noting that $\tilde{\eta}=x/\left(2\sin^2{\theta^*}\right)+\mathcal{O}\left(x^2\right)$. The resulting expansion of Eq.~\eqref{eq:DefCOM} exactly reproduces Eq.~\eqref{eq:fixed_order_COM} at $\eta=0$. (This computation is analogous to the derivation in Appendix~\ref{app:Derivation_zphi_fixedorder} In these kinematics the counterpart of the inequality in Eq.~\eqref{eq:kinematic_region} is given by $4z_\phi \cos^2\theta^* \ll 1$.) Since the $z_\phi$ structure remains the same after the change of variables one can still formulate a one-dimensional recoil factorization formula in the parent-boson center-of-mass frame by keeping it differential in $\sin^2\theta^*$ instead of $z_\theta$. 
\section{Phenomenological Studies}
\label{sec:Phenomenology}

To illustrate the factorization formulas we make phenomenological predictions and contrast them with \textsc{Pythia} simulations. In Section~\ref{sec:InclusiveEEC} we make predictions for the fully inclusive EEC following~Eq.~\eqref{eq:epem_EEC} for an asymmetric $e^+e^-$ detector such as the BaBar detector for which the intermediate virtual photon is boosted in the lab frame. In this fully inclusive EEC the Collins effect is suppressed~\cite{Kang:2023big} (see also Eq.~\eqref{eq:CollinsJetfun}); hence we neglect it and do we do not study any azimuthal modulations. Instead we test the perturbative convergence of the factorization formula when compared with \textsc{Pythia} simulations~\cite{Bierlich:2022pfr}. 

In Section~\ref{sec:Zdecaypheno} we consider hadronic Z-decay in the reaction $e^+e^- \rightarrow Z\gamma$. The required Z-boson density matrix for this observable can be computed from the helicity amplitudes given in~\cite{Ellis:2019zex}. In this setting we will not sum over all possible hadrons in the final state to allow for a non-zero Collins-type jet function. We also find non-vanishing azimuthal modulation $\cos{2\left(\bar{\phi}-\varphi\right)}$ which allows us to search for the Collins effect by defining charged particle pair ratios.

\subsection{(Inclusive) EEC in asymmetric $e^+e^-$ collisions}
\label{sec:InclusiveEEC}
Including TMD evolution the relevant factorization formula for the unpolarized term reads
{\allowdisplaybreaks
\begin{align}
W^{\mathrm{U}}_{\gamma\gamma}
&=N_c\sum_q e_q^2\,H(\mu_H,Q)\,
\frac{Q^4p_T}{16\pi^3}\,
\frac{\mathcal{K}(z_\theta;\eta)}{\sqrt{z_\phi}}
\notag\\
&\quad\times\int_0^\infty db\,
\cos\!\big(2bp_T\sqrt{z_\phi}\big)\,
S_{n_1n_2}(b,\mu_S,\nu_S)\,
J_q\big(b,\mu_J,\zeta_1/\nu_J^2\big)\,
J_{\bar q}\big(b,\mu_J,\zeta_2/\nu_J^2\big)
\notag\\
&\quad\times \mathrm{exp}\Bigg[
\int_{\mu_H}^{\mu}\frac{d\mu'}{\mu'}\,
\gamma_H^q(Q,\mu')
+
\int_{\mu_J}^{\mu}\frac{d\mu'}{\mu'}\,
\widetilde{\gamma}_J^q
\left(\mu',\frac{\zeta_1}{\nu_J^2}\right)
\notag\\
&\qquad\qquad+
\int_{\mu_J}^{\mu}\frac{d\mu'}{\mu'}\,
\widetilde{\gamma}_J^q
\left(\mu',\frac{\zeta_2}{\nu_J^2}\right)
+
\int_{\mu_S}^{\mu}\frac{d\mu'}{\mu'}\,
\widetilde{\gamma}_{S,n_1n_2}^q(\mu',\nu_S)
-2S_{\mathrm{NP}}(b)
\Bigg]
\notag\\
&\quad\times
\left(\frac{\nu_J}{\nu_S}\right)^{
\widetilde{\gamma}_\nu^q(b,\mu)-4D_{\mathrm{NP}}(b)}.
\label{eq:HUnpol_evolved}
\end{align}
}
where $S_{\mathrm{NP}}\left(b\right)$ and $D_{\mathrm{NP}}\left(b\right)$ encode the non-perturbative parts of the Sudakov exponential and the Collins-Soper kernel respectively. We follow the anomalous dimension definitions from~\cite{Ebert:2020sfi}
{\allowdisplaybreaks
\begin{align}
\gamma_H^q(Q,\mu)
&\equiv \frac{\partial\ln H}{\partial\ln\mu}
=4\Gamma^q\ln\frac{Q}{\mu}+4\gamma^q,
\\
\widetilde{\gamma}_J^q\left(\mu,\frac{\zeta_1}{\nu_J^2}\right)
&\equiv \frac{\partial\ln J_q}{\partial\ln\mu}
=-\Gamma^q\ln\frac{\zeta_1}{\nu_J^2}
-2\gamma^q+\gamma_s^q,
\\
\widetilde{\gamma}_J^q\left(\mu,\frac{\zeta_2}{\nu_J^2}\right)
&\equiv \frac{\partial\ln J_{\bar q}}{\partial\ln\mu}
=-\Gamma^q\ln\frac{\zeta_2}{\nu_J^2}
-2\gamma^q+\gamma_s^q,
\\
\widetilde{\gamma}_{S,n_1n_2}^q(\mu,\nu_S)
&\equiv \frac{\partial\ln S_{n_1n_2}}{\partial\ln\mu}
=4\Gamma^q\ln\frac{\mu}{\nu_S\sqrt{z_\theta}}
-2\gamma_s^q.
\end{align}
}
Here $\sqrt{\zeta_i}=2E_i$ and $\Gamma^q=\Gamma_{\mathrm{cusp}}^q(\alpha_s(\mu))$. The noncusp terms $\gamma^q$ and $\gamma_s^q$ are also evaluated at $\alpha_s(\mu)$. In the last equation we used 
\begin{equation}
\widetilde{\gamma}_{S,n_1n_2}^q(\mu',\nu_S)
=
\widetilde{\gamma}_{S,n\bar n}^q
\left(\mu',\nu_S\sqrt{z_\theta}\right),
\end{equation}
i.e. the $\mu$-anomalous dimension of the soft function for the directions $n_1,n_2$ is related to that of the standard back-to-back TMD soft function by a geometric rescaling of the rapidity scale~\cite{Gao:2023ivm}.
The soft rapidity anomalous dimension is defined by
\begin{equation}
\widetilde{\gamma}_\nu^q(b,\mu)
\equiv \frac{\partial\ln S_{n_1n_2}}{\partial\ln\nu}.
\end{equation}
Its perturbative part in the $b_*$ prescription~\cite{Collins:1984kg} is
\begin{equation}
\widetilde{\gamma}_{\nu,\mathrm{pert}}^q(b_*,\mu)
=
-4\int_{\mu_{b_*}}^{\mu}\frac{d\mu'}{\mu'}\,
\Gamma_{\mathrm{cusp}}^q\!\left(\alpha_s(\mu')\right)
+2\gamma_r^q\!\left(\alpha_s(\mu_{b_*})\right),
\qquad
\mu_{b_*}=\frac{b_0}{b_*}.
\end{equation}
where $b^* = b/\sqrt{1+b^2/b^2_{\mathrm{max}}}$  with  $b_0 =2e^{-\gamma_E}$ and $b_{\mathrm{max}}=\,1\, \mathrm{GeV}^{-1}$. We use the small-$b$ OPE for the unpolarized quark TMDFF 
\begin{equation}
     D_{h/i}\left(\xi,b,\mu_J,\frac{\zeta_i}{\nu_J^2}\right) = \frac{1}{\xi}\sum_j \int_\xi^1 \frac{d\tau}{\tau}\, C_{j/i}\left(\frac{b}{\tau},\tau,\mu_J,\frac{\zeta_i}{\nu_J^2}\right)D_{h/j}\left(\frac{\xi}{\tau},\mu_J\right)+\mathcal O\!\left(b^2\Lambda_{\rm QCD}^2\right),
\end{equation}
from which it follows that the quark EEC jet function can be written as
\begin{align}
J_i\!\left(b,\mu_J,\frac{\zeta_i}{\nu_J^2}\right)={}&\sum_j\int_0^1 d\tau\,\tau\,C_{j/i}\!\left(\frac{b}{\tau},\tau,\mu_J,\frac{\zeta_i}{\nu_J^2}\right)\sum_{h}\int_0^1 d\xi\,\xi\,D_{h/j}(\xi,\mu_J)+\mathcal O\!\left(b^2\Lambda_{\rm QCD}^2\right)\, . \label{eq:UnpolQuarkJetfunction}
\end{align}
We drop the corrections to the OPE and keep the jet function as purely perturbatively calculable moments of the Wilson coefficients, since the momentum sum rule of Eq.~\eqref{eq:SumRuleUnpol} removes the dependence on the collinear functions. All the non-perturbative effects are kept only in the non-perturbative Sudakov and the non-perturbative Collins-Soper kernel in Eq.~\eqref{eq:HUnpol_evolved}. For the Collins function we have an analogous expression
\begin{equation}
\begin{split}
W^{\perp}_{\gamma\gamma}
&=N_c\sum_q e_q^2\,H(\mu_H,Q)\,
\frac{Q^4p_T}{16\pi^3}\,
\frac{\mathcal{K}(z_\theta;\eta)}{\sqrt{z_\phi}}
\\
&\quad\times\frac{1}{4}\int_0^\infty db\,b^2\,
\cos\!\big(2bp_T\sqrt{z_\phi}\big)\,
S_{n_1n_2}(b,\mu_S,\nu_S)\,
J_q^\perp\big(b,\mu_J,\zeta_1/\nu_J^2\big)\,
J_{\bar q}^\perp\big(b,\mu_J,\zeta_2/\nu_J^2\big)
\\
&\quad\times\exp\Bigg[
\int_{\mu_H}^{\mu}\frac{d\mu'}{\mu'}\,
\gamma_H^q(Q,\mu')
+
\int_{\mu_J}^{\mu}\frac{d\mu'}{\mu'}\,
\widetilde{\gamma}_J^q
\left(\mu',\frac{\zeta_1}{\nu_J^2}\right)
\\
&\qquad\qquad+
\int_{\mu_J}^{\mu}\frac{d\mu'}{\mu'}\,
\widetilde{\gamma}_J^q
\left(\mu',\frac{\zeta_2}{\nu_J^2}\right)
+
\int_{\mu_S}^{\mu}\frac{d\mu'}{\mu'}\,
\widetilde{\gamma}_{S,n_1n_2}^q(\mu',\nu_S)
-2S_{\mathrm{NP}}^{\mathrm{Collins}}(b)
\Bigg]
\\
&\quad\times
\left(\frac{\nu_J}{\nu_S}\right)^{
\widetilde{\gamma}_\nu^q(b,\mu)-4D_{\mathrm{NP}}(b)},
\label{eq:HCollins_evolved}
\end{split}
\end{equation}
where $S^{\mathrm{Collins}}_{\mathrm{NP}}\left(b\right)$ encodes the non-perturbative part of the Sudakov exponential for the Collins function, while the non-perturbative Collins-Soper kernel $D_{\mathrm{NP}}\left(b\right)$ is the same as in Eq.~\eqref{eq:HUnpol_evolved} since it does not depend on spin. The small-$b$ OPE for the Collins function is given by~\cite{Kang:2015msa}
\begin{align}
\tilde{H}^{\perp}_{h/i}\left(\xi,b,\mu_J,\frac{\zeta_i}{\nu_J^2}\right) ={}&\frac{1}{\xi}\sum_j \int_\xi^1 \frac{d\tau}{\tau}\, \delta C_{j/i}^{\mathrm{Collins}}\left(\frac{b}{\tau},\tau,\mu_J,\frac{\zeta_i}{\nu_J^2}\right)\hat{H}^{(3)}_{h/j}\left(\frac{\xi}{\tau},\mu_J\right)\nonumber\\
&+\left[A_{j/i} \,\tilde{\otimes}\,\hat{H}_{F,h/j}\right]\left(\xi,b,\mu_J,\frac{\zeta_i}{\nu_J^2}\right)+\mathcal O\!\left(b^2\Lambda_{\rm QCD}^2\right).\label{eq:CollinsOPE}
\end{align}
The second term only starts at an order higher in $\alpha_s$ relative to the first one and we will therefore neglect it in this analysis. Then the Collins-type EEC jet function is given by
\begin{equation}
\begin{split}
    J^\perp_{i}\left(b,\mu_J,\frac{\zeta_i}{\nu_J^2}\right)  ={}& \sum_j \int_0^1 d\tau\, \tau \, \delta C^{\mathrm{Collins}}_{j/i}\left(\frac{b}{\tau},\tau,\mu_J,\frac{\zeta_i}{\nu_J^2}\right)\,\sum_{h}\int_0^1 d\xi\, \xi \, \hat{H}^{(3)}_{h/j}\left(\xi,\mu_J\right)\\
    &+\mathcal O\!\left(b^2\Lambda_{\rm QCD}^2\right).
\end{split} \label{eq:CollinsJetfun}
\end{equation}
The Schäfer-Teryaev (ST) sum rule~\cite{Schafer:1999kn,Meissner:2010cc} (Eq.~\eqref{eq:CollinsSumRule}) makes this Collins-type EEC jet function vanish up to the neglected corrections. Therefore for this fully inclusive EEC we will only retain the unpolarized prediction.

To make phenomenological predictions we choose $\mu_H=\nu_J=Q$
and $\mu_J=\mu_S=\nu_S=b_0/b^*$.  The two-loop soft and hard functions can be found in~\cite{Moult:2018jzp} and the two-loop unpolarized quark EEC jet functions in~\cite{Ebert:2020sfi}.  The cusp anomalous dimensions up to four loops were computed in \cite{Moch:2018wjh,Moch:2017uml,Davies:2016jie,Henn:2019swt,Korchemsky:1987wg,Moch:2004pa} and the rapidity anomalous dimension was given in~\cite{Li:2016ctv} to three loops. The quark anomalous dimension has been computed up to three loops in \cite{Moch:2005id,Becher:2006mr}. The soft anomalous dimension has also been computed to three loops in~\cite{Li:2014afw}. The QCD beta function up to four loops was given in~\cite{Tarasov:1980au,Larin:1993tp,vanRitbergen:1997va,Czakon:2004bu}. For the non-perturbative pieces in Eq.~\eqref{eq:HUnpol_evolved} we took the non-perturbative Sudakov exponential and Collins-Soper kernel from~\cite{Kang:2024dja}. In particular we took their $D_{\mathrm{NP}}^{\mathrm{Fit 1}}(b)$ model.

\begin{figure}[t]
  \centering
  \includegraphics[width=1\textwidth]{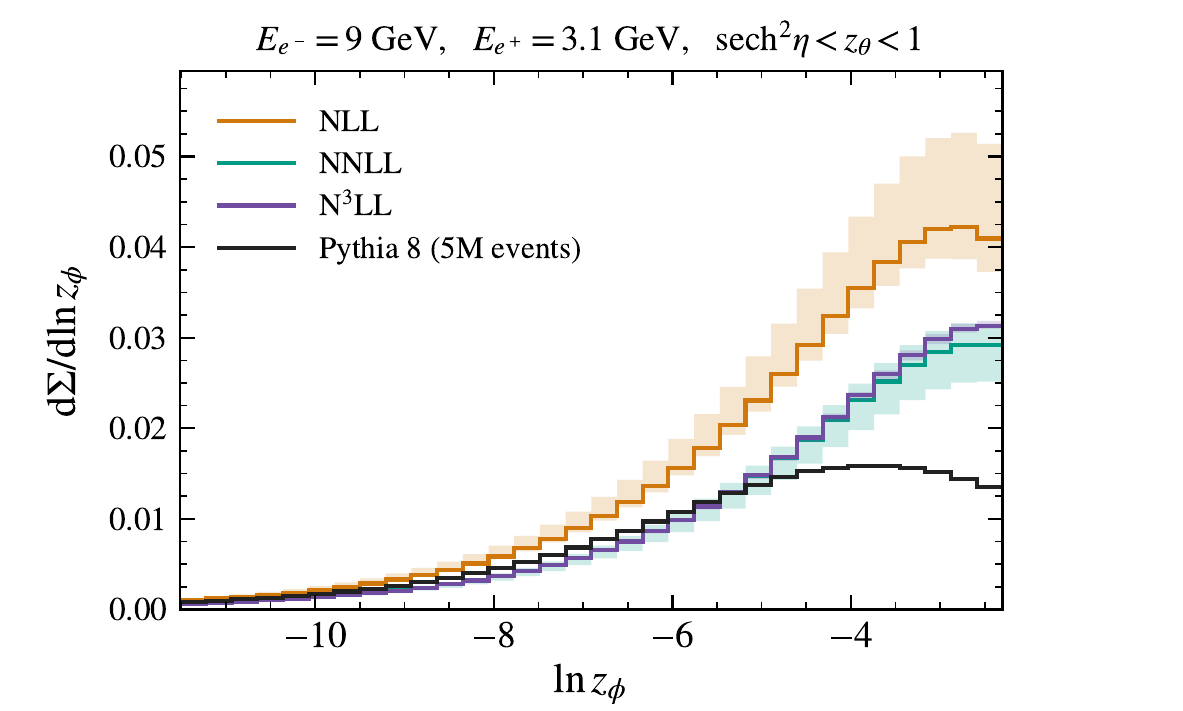}\\[2pt]
  \makebox[0.15\textwidth]{}\makebox[0.15\textwidth]{}
  \caption{Inclusive EEC for $e^+e^- \rightarrow\mathrm{hadrons}$ in the lab frame with an asymmetric beam which produces virtual photons at a rapidity $\eta =0.53 $. 
  Theoretical uncertainty bands are generated by varying the initial scales $\mu_H$ and $\nu_J$ by a factor of 2. The simulation curve was obtained from 5 million accepted events.}
  \label{fig:EEC_Inclusive}
\end{figure}

We choose kinematics relevant to the BaBar detector, that is $E_{e^-} =9.0 \, \mathrm{GeV}$ and $E_{e^+} = 3.1 \, \mathrm{GeV}$ which annihilate into a virtual photon with rapidity $\eta = 0.53$. In Fig.~\ref{fig:EEC_Inclusive} we integrate over the allowed $z_\theta$ phase space and compare the theoretical curves with 5 million \textsc{Pythia} events with hadron decay disabled and $n_f=5$. The theoretical uncertainty bands are produced by varying $\mu_H$ and $\nu_J$ by a factor of 2. We observe that below $z_\phi \sim 10^{-2} $ the N3LL theoretical curve agrees well with the \textsc{Pythia} simulation. Besides the dependence on the non-perturbative fitted input, small deviations from the theoretical curves are expected since the $z_\theta \rightarrow1$ endpoint brings about $p_T \sim \Lambda_{\mathrm{QCD}}$ corrections to the factorization formula (see discussion around Eq.~\eqref{eq:phi_LP}) and the $z_\theta \rightarrow\sech^2\eta$  endpoint also violates the approximations described in Eq.~\eqref{eq:kinematic_region}.
\subsection{Photon-tagged hadronic $Z$ decay (non-inclusive) EEC in $e^+e^-\rightarrow Z\gamma$}
  \label{sec:Zdecaypheno}
  
We now explore the EEC measured on the hadronic decay products of a $Z$ boson produced in the process $e^+e^- \rightarrow Z\gamma$ in the $e^+e^-$ center-of-mass frame. We aim to verify whether such an EEC can be used to probe the Collins effect. Recall that the possible angular modulations in the factorization formula in Eq.~\eqref{eq:Z_fact} need as an input the produced $Z$ boson density matrix. We computed it from the tree-lelve helicity amplitudes in~\cite{Ellis:2019zex} by summing over the lepton helicities (since we consider an unpolarized beam) and tracing over the photon ones. We carried this out explicitly in Appendix~\ref{app:ZDensityMatrix}. With those inputs we find that the factorization formula is given by
\begin{equation}
\begin{aligned}
\frac{d\Sigma_Z}{dz_\phi\,dz_\theta\,d\bar\phi}
&= \frac{\operatorname{sech}^2\eta}{2M_Z^3\Gamma_{Z,0}}
\Biggl\{
\left[
\frac{1+\cos^2\theta^*}{2}
\bigl[\rho_{++}(\theta_\gamma)+\rho_{--}(\theta_\gamma)\bigr]
+\sin^2\theta^*\,\rho_{00}(\theta_\gamma)
\right]
W_{ZZ}^{\mathrm{U}}
\\
&\quad
+\sin^2\theta^*\left[
\frac{1}{2}
\bigl[\rho_{++}(\theta_\gamma)+\rho_{--}(\theta_\gamma)\bigr]
-\rho_{00}(\theta_\gamma)
\right]
W_{ZZ}^{\perp}
\\
&\quad
-\rho_{+-}(\theta_\gamma)\cos\!\bigl[2(\bar\phi-\varphi)\bigr]
\left[
\sin^2\theta^*\,W_{ZZ}^{\mathrm{U}}
+(1+\cos^2\theta^*)\,W_{ZZ}^{\perp}
\right]
\Biggr\}.
\end{aligned}
\label{eq:Z_EECFact}
\end{equation}
\begin{figure}[t]
  \centering
  \includegraphics[width=1.0\textwidth]{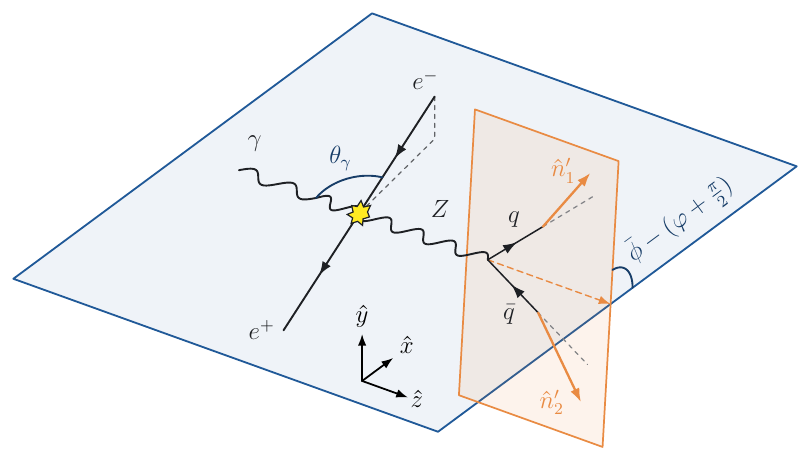}\\[2pt]
  \makebox[0.15\textwidth]{}\makebox[0.15\textwidth]{}
  \caption{Underlying process for the photon-tagged $Z$-boson hadronic decay EEC in the $e^+e^-$ center-of-mass frame. Note that this illustration assumes $z_\phi\ll1$ kinematics, in which the $Z$ boson and the two detectors are approximately coplanar (compare with Fig.~\ref{fig:EEC_Boosted}).
  The $\hat{z}$ axis is always chosen to be aligned with the momentum of the $Z$ boson, while the transverse directions are arbitrary until $\varphi$ is fixed.  In this illustration we chose to align the $\hat{x}$ axis opposite to the direction of the incoming electron, which then gives $\varphi = 0$ following the convention in Appendix~\ref{app:ZDensityMatrix}.}
  \label{fig:Planes}
\end{figure}
The $Z$-boson density matrix $\rho'(\theta_\gamma,\theta^*,\bar{\phi}-\varphi)$ appearing in Eq.~\eqref{eq:Z_fact} in the quark-helicity frame, has been rewritten in terms of the production density matrix $\rho(\theta_\gamma)$ computed in Eq.~\eqref{eq:Zdensitymatrix}, whose quantization axis follows the $Z$ flight direction in the $e^+e^-$ COM frame. Here $\theta_\gamma$ is the angle between the photon and the incoming electron in that frame, $\sin{^2\theta^*}$ is defined in Eq.~\eqref{eq:sinCOM} and $\bar{\phi}$ is the average azimuth of the correlated hadron pair about the $Z$ flight axis. The reference azimuth $\varphi$ specifies the direction opposite to the incoming electron momentum projected onto the plane perpendicular to the $Z$ momentum, as defined in Eq.~\eqref{eq:varphi_def}. These angles are illustrated in Fig.~\ref{fig:Planes}. 

\begin{figure}[t]
  \centering
  \includegraphics[width=0.95\textwidth]{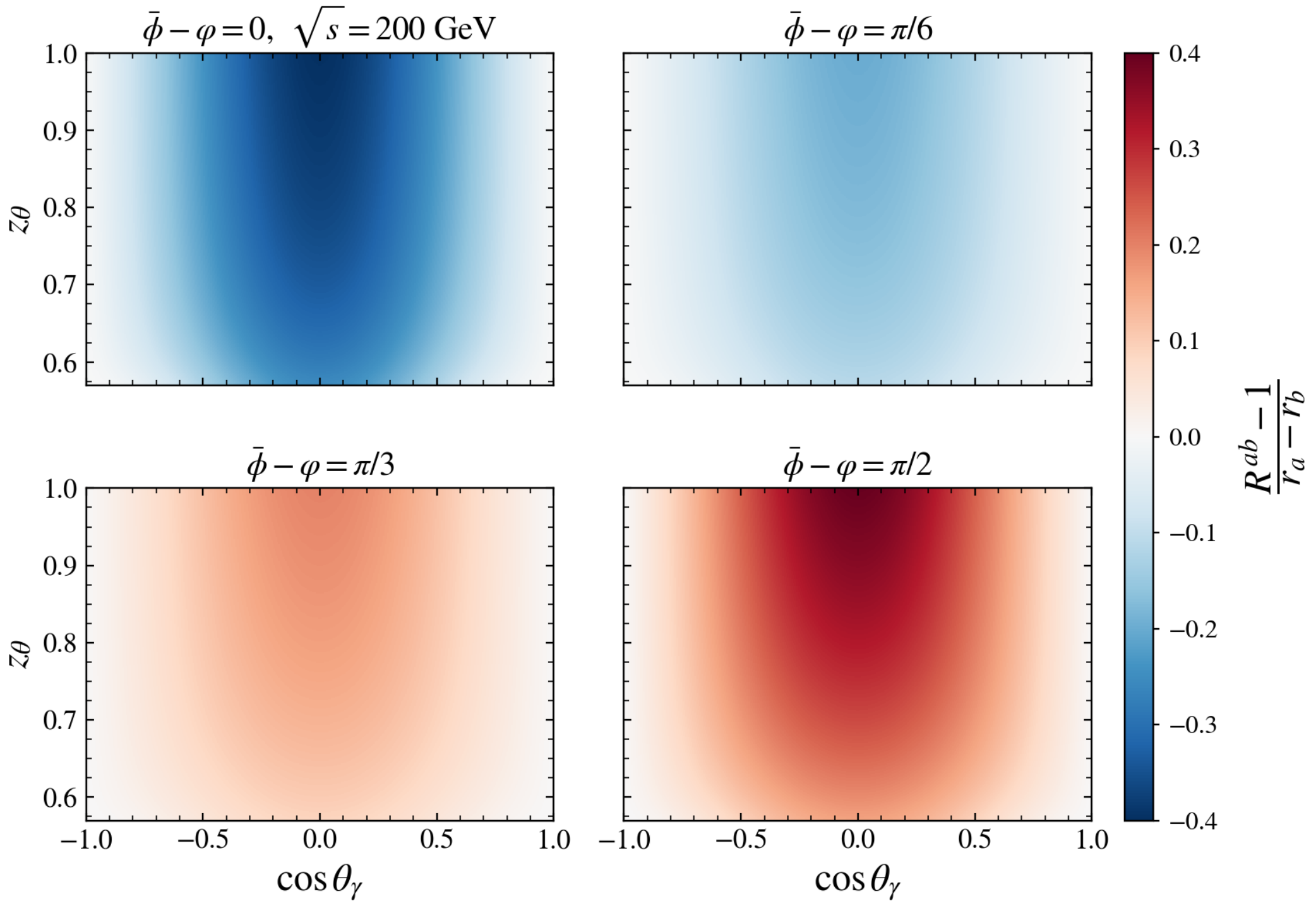}\\[2pt]
  \makebox[0.15\textwidth]{}\makebox[0.15\textwidth]{}
  \caption{Expected angular dependence of the leading-order expansion of $R^{ab}$ assuming $r_a\sim r_b\ll1$ at $\sqrt{s} = 200\, \mathrm{GeV}$. To increase sensitivity to the Collins effect we look for kinematics that maximize the variation as a function of $\bar{\phi}-\varphi$.}
  \label{fig:Coefficient_Ratio}
\end{figure}

Even though in Eq.~\eqref{eq:Z_EECFact} the azimuthal $\cos{2\left(\bar{\phi}-\varphi\right)}$ multiplies both the unpolarized and Collins-dependent terms, we can still define a charge-dependent double ratio that isolates the Collins effect following~\cite{Belle:2005dmx,Belle:2008fdv,BaBar:2013jdt}. First we introduce the notation 
\begin{equation}
    r_{\mathbb{S}} \equiv \frac{ W_{ZZ}^{\perp,\,\mathbb{S}} }{W_{ZZ}^{\mathrm{U},\,\mathbb{S}} },
\end{equation}
where $\mathbb{S}$ denotes that the unpolarized or Collins-dependent terms are not computed with an inclusive EEC, but rather summed only over a subset of hadrons. Then the EEC on a subset of hadrons $\mathbb{S} $ normalized by its $\bar{\phi}$ average is given by
\begin{equation}
\Sigma_{Z,\mathbb{S}} \frac{1}{\langle \Sigma_{Z,\mathbb{S}}\rangle}_{\bar{\phi}} = 1-\rho_{+-}\left(\theta_\gamma\right) \frac{\sin^2\theta^*+\left(1+\cos^2{\theta^*}\right)r_{\mathbb{S}}}{1-\rho_{00}\left(\theta_\gamma\right)
-\frac{1-3\rho_{00}\left(\theta_\gamma\right)}{2}\sin^2\theta^* \left(1-r_{\mathbb{S}}\right)}\cos{2\left(\bar{\phi}-\varphi\right)},
\end{equation}
where we have used $\rho_{++}\left(\theta_\gamma\right)+\rho_{--} \left(\theta_\gamma\right)+\rho_{00}\left(\theta_\gamma\right)= 1$.
Constructing the same operation for a different subset of hadrons we define the double ratio
\begin{equation}
    R^{ab} = \frac{1-\rho_{+-}\left(\theta_\gamma\right)\cos{2\left(\bar{\phi}-\varphi\right)} \frac{\sin^2\theta^*+\left(1+\cos^2{\theta^*}\right)r_{a}}{1-\rho_{00}\left(\theta_\gamma\right)
-\frac{1-3\rho_{00}\left(\theta_\gamma\right)}{2}\sin^2\theta^* \left(1-r_{{a}}\right)}}{1-\rho_{+-}\left(\theta_\gamma\right) \cos{2\left(\bar{\phi}-\varphi\right)}\frac{\sin^2\theta^*+\left(1+\cos^2{\theta^*}\right)r_{b}}{1-\rho_{00}\left(\theta_\gamma\right)
-\frac{1-3\rho_{00}\left(\theta_\gamma\right)}{2}\sin^2\theta^* \left(1-r_{b}\right)}}. \label{eq:ColinsRatio}
\end{equation}
We expect the Collins-dependent terms to be sub-leading to the unpolarized terms. Then taking $r_a\sim r_b \ll1$ 
\begin{equation}
\begin{split}
R^{ab}\simeq{}&1
-(r_a-r_b)\,
\frac{\rho_{+-}(\theta_\gamma)\cos\!\bigl[2(\bar{\phi}-\varphi)\bigr]}
     {1-\rho_{00}(\theta_\gamma)
      -\frac{1-3\rho_{00}(\theta_\gamma)}{2}\sin^2\theta^*}
\\
&\times
\frac{2\bigl[(1-\rho_{00}(\theta_\gamma))\cos^2\theta^*
             +\rho_{00}(\theta_\gamma)\sin^2\theta^*\bigr]}
     {1-\rho_{00}(\theta_\gamma)
      -\frac{1-3\rho_{00}(\theta_\gamma)}{2}\sin^2\theta^*
      -\rho_{+-}(\theta_\gamma)\sin^2\theta^*
       \cos\!\bigl[2(\bar{\phi}-\varphi)\bigr]}, \label{eq:Ration_Expanded}
\end{split}
\end{equation}
Within the leading-power factorization at fixed kinematics, the double ratio removes the common unpolarized angular factor and probes differences in the relative Collins contributions. The size of modulation, which is determined by $r_a$ and $r_b$, can be obtained by fitting an experimental curve to Eq.~\eqref{eq:ColinsRatio} at some fixed kinematics, although finite-bin and acceptance effects must be controlled when applying this cancellation to data. To select kinematics that maximize the size of the azimuthal modulation in the double ratio, we plotted the angular coefficient of $(R^{ab}-1)/\left(r_a-r_b\right)$ in Eq.~\eqref{eq:Ration_Expanded}  at $\sqrt{s}=200 \, \mathrm{GeV}$, an energy within the range covered by LEP2, whose archived data are currently being reanalyzed~\cite{Electron-positronAlliance:2026ile}, and comparable to that of proposed $e^+e^-$ Higgs factories. The results are shown in Figure~\ref{fig:Coefficient_Ratio}. From these plots it is clear that in a range $|\cos{\theta_\gamma}|<0.5$ the Collins modulation is expected to be the largest.
Following the charge combinations used by BaBar~\cite{BaBar:2013jdt},
we restrict the analysis to charged-pion pairs and denote unlike-sign
pairs by $U$, like-sign pairs by $L$, and all charged pairs by $C=U+L$.
We focus on $R^{UL}$, whose leading modulation probes $r_U-r_L$, since at fixed kinematics it is typically larger than $R^{UC}$. The restriction to charged pions also changes the fragmentation
input. The sum rules used to simplify the fully inclusive EEC in the
previous section do not apply to these charge-selected samples.
We therefore retain the full unpolarized and Collins TMD
fragmentation functions in the prediction for $R^{UL}$ and compare
the resulting azimuthal distribution with the event-generator
simulations.

On the simulation side we use the package \textsc{StringSpinner}~\cite{Kerbizi:2026iws,Kerbizi:2021pzn,Kerbizi:2023luv,Kerbizi:2024vpd} which incorporates quark spin effects in the \textsc{Pythia} event generator using the string$+^3P_0$ model of hadronization~\cite{Kerbizi:2018qpp}. Spin correlations between the produced quark-antiquark pair are propagated along the string fragmentation chain by using the rules of the model and are implemented for the production of pseudoscalar and vector mesons. \textsc{StringSpinner} takes as input the spin correlation matrix $C_{ij} = \mathrm{Tr}\left[\rho_{q\bar{q}}\left(\sigma_i \otimes \sigma_j\right)\right]$ which is computed from the quark-antiquark density matrix with the helicities quantized in the quark helicity frame. We construct it in Appendix~\ref{app:qqbarDensityMatrix}.  For the free parameters of the algorithm, we use the preliminary tuning quoted in~\cite{Kerbizi:2026iws}. We contrast these results with theoretical predictions using the fitted Collins TMDFF and approximate NLL accuracy as defined in~\cite{Kang:2015msa}. For this study we therefore adopt their conventions for both TMDFFs, running coupling and non-perturbative models which involve the DSS14 NLO collinear FF set~\cite{deFlorian:2014xna}. 

\begin{figure}[t]
  \centering
  \includegraphics[width=1.0\textwidth]{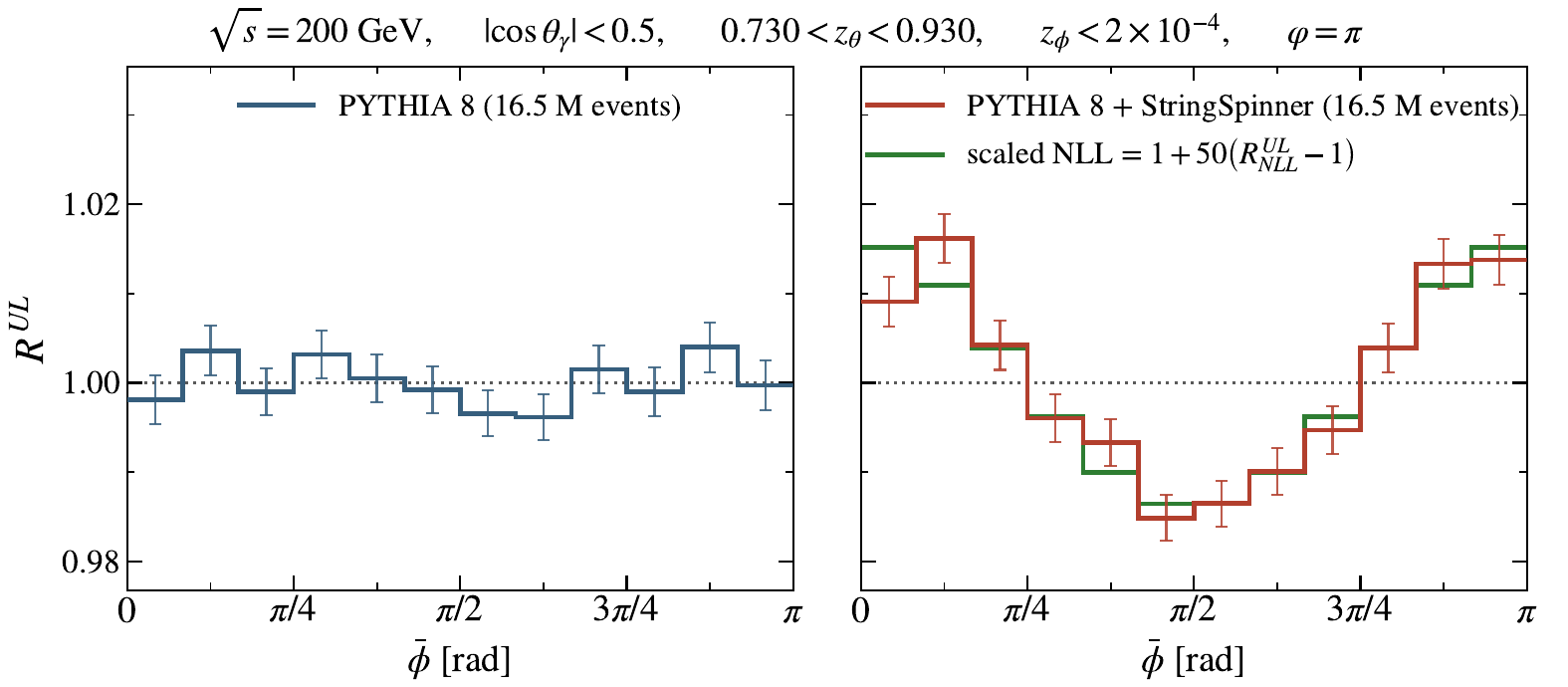}\\[2pt]
  \makebox[0.15\textwidth]{}\makebox[0.15\textwidth]{}
  \caption{Left panel: Ratio of EEC constructed from pairs of unlike-charge pions to EEC constructed from like-charge pions computed with plain \textsc{Pythia}~8. Right panel: Same ratio,  with Collins modulation incorporated. To compute the red histogram the fragmentation algorithm of \textsc{Pythia}~8 is supplemented by the \textsc{StringSpinner}~\cite{Kerbizi:2026iws,Kerbizi:2021pzn,Kerbizi:2023luv,Kerbizi:2024vpd} package which models the Collins effect. The green curve is computed by using the fitted Collins function from~\cite{Kang:2015msa} directly in the factorization formula and scaling its magnitude around 1 by a factor of 50. We used their approximate NLL conventions here. 
 The simulation uncertainties are the standard deviation over 5,000 bootstrap replicas, obtained by resampling blocks of 5,000 generated events with replacement. Both panels display simulations with 16.5 million accepted events.}   \label{fig:RatioPlots}
\end{figure}

Based on the scan of kinematics in Fig.~\ref{fig:Coefficient_Ratio} we select the box $|\cos{\theta_\gamma|<0.5}$ and $0.730<z_\theta<0.930$ where the oscillations as a function of $\bar{\phi}-\varphi$ are the largest. We avoid the $z_\theta$ endpoint since then the hierarchy $p_T\gg \Lambda_{\mathrm{QCD}}$ would be violated for arbitrarily small $\left(1-z_\theta\right)$  (see discussion around Eq.~\eqref{eq:phi_LP}). We also bin in $z_\phi <2\times 10^{-4}$ to ensure we are well within the region of validity of the factorization formula. The results are shown in Fig.~\ref{fig:RatioPlots}. Here we worked with the three lightest quarks on both theory and simulation and we disabled hadron decay on the simulation side. The left panel shows that when plain \textsc{Pythia} is used to simulate the events the ratio is consistent with one, which is expected since according to Eq.~\eqref{eq:Ration_Expanded} deviations from unity are due to the Collins effect. On the right panel we see that once \textsc{StringSpinner} is turned on the ratio shows a $\sim 1 \%$ modulation. When contrasted with the NLL theoretical expectation, we see that the theory has a modulation about unity which is $\sim 50$ times smaller. This disagreement on the size of the modulation may be due to the different models used to construct the ratios $r_U$ and $r_L$ which are determined by the non-perturbative parameters fitted in~\cite{Kerbizi:2026iws} for \textsc{StringSpinner} and in~\cite{Kang:2015msa} for the Collins function. Nevertheless, since the shapes of the two distributions agree fairly well, we interpret this as evidence of the azimuthal modulation derived in Eq.~\eqref{eq:Z_EECFact}. Statistical uncertainties for the simulation data are obtained from 5,000 bootstrap replicas, resampling blocks of 5,000 generated events with replacement. Both charge channels and all azimuthal bins are resampled jointly, and the individually normalized double ratio is recomputed in each replica. The error bars show the standard deviation of these replicas. 

\section{Conclusions and outlook}
\label{sec:Conclusions}
Through studying the transformation properties of the energy flow operator we decomposed the EEC hadronic tensor into a few independent functions, for flavor-neutral and flavor-changing axial and vector currents building on the work in~\cite{Riembau:2025isw}. For flavor-changing currents, time reversal shows that the only structure that differs between in- and out-detectors is the symmetric, parity-odd function $W_S$. In perturbation theory it arises from the imaginary part of loop diagrams and requires a mass difference between the quark and the antiquark, while non-perturbatively it could also receive contributions from rescattering during hadronization. Both $W_S$ and the antisymmetric structure $W_A$ vanish for massless quarks, which is also the regime in which the energy flow operator transforms simply under boosts. We therefore restricted ourselves to massless quarks and boosted the remaining functions $W_{0,1,2}$, together with the scalar EECs for Higgs decay. By studying their endpoint enhancement, we identified that in the small-acoplanarity limit boosted EECs display a one-dimensional-recoil TMD factorization inherited from the well-known two-dimensional-recoil factorization in the back-to-back limit of the EEC in the COM frame. The observable we propose has the simple geometric requirement that the parent boosted boson and the pair of detected hadrons approximately lie in the same plane.

We then constructed the factorization formula in $e^+e^-$ colliders and for hadronically decaying electroweak and Higgs bosons, and we verified that it correctly matches the real-emission contribution to the fixed-order results to one loop. Keeping a completely inclusive EEC we made phenomenological predictions at the kinematics of the BaBar experiment by taking from the literature the perturbative computations of the hard, jet and soft functions, and their corresponding anomalous dimensions, up to N3LL accuracy. The theoretical curves showed the expected perturbative convergence for small acoplanarities $z_\phi< 10^{-2}$ and comparison with \textsc{Pythia} simulations showed good agreement. To test the feasibility of probing the Collins effect with our formalism we considered photon-tagged hadronic $Z$ decays in $e^+e^- \rightarrow Z\gamma$, in which we restricted the correlated hadrons to charged pions and constructed the unlike-sign to like-sign double ratio $R^{UL}$. Within the leading-power factorization formula, this double ratio cancels the common unpolarized angular dependence and probes differences between the relative Collins contributions in the two charge channels. The resulting modulation depends on the orientation of the detected pair about the $Z$-boson momentum. We compared theoretical predictions at NLL based on extractions of the Collins function with \textsc{Pythia} + \textsc{StringSpinner} simulations and found agreement in the shape of the modulations, but a significantly larger modulation magnitude on the simulation side.

Our formalism thus opens a new avenue for studying energy--energy correlations in boosted hadronic systems and probing spin-dependent fragmentation, with sensitivity to the quark Collins function and potential applications to the fragmentation of linearly polarized gluons. A natural extension is its application to the hadronic decays of heavy particles produced at the LHC. In contrast with the TEEC, whose azimuthal correlations are measured about the beam axis and whose factorization involves beam functions and soft radiation connecting the initial and final states, our observable is defined entirely with respect to the momentum of the decaying boson, which is reconstructed event by event from its decay products. For a color-singlet resonance in the narrow-width approximation, the factorization formulas derived here therefore apply independently of the production mechanism, which enters only through the density matrix of the boson and the distribution of its boost. Another promising direction is to investigate the small-acoplanarity limit in boosted heavy-quarkonium decays, combining NRQCD~\cite{Bodwin:1994jh} with soft-collinear factorization. In particular, the leading color-singlet contribution to $\eta_b$ decay proceeds through two gluons~\cite{Feng2017Pseudoscalar}, suggesting a factorization structure analogous to that developed here for Higgs decay. Varying the measured transverse-recoil scale relative to the quarkonium bound-state scales could perhaps provide sensitivity to interesting non-perturbative bound-state dynamics.

\section*{Acknowledgments}
We  thank Kyle Lee and Jani Penttala for discussions and comments. This work is supported by the National Science Foundation under grant No.~PHY-2515057. 

\appendix
\section{Transformation properties of the energy flow operator}

\subsection{Discrete transformations}
In this Appendix we discuss the C, P and T transformation properties of the energy flow operator defined in~\eqref{eq:energy_flow_matrix_element_com}. These properties are used in the main text to derive the tensor decomposition of the EEC hadronic tensor in~\eqref{eq:eec-tensor-def}.  Since nothing in Eq.~\eqref{eq:energy_flow_matrix_element_com} depends on charge, we immediately get
\begin{equation}
    \mathcal{C} \mathcal{E}_{\mathrm{out}}\left(n\right)\mathcal{C}^{-1} = \mathcal{E}_{\mathrm{out}}\left(n\right).
\end{equation}
Let us now conjugate the energy flow operator in Eq.~\eqref{eq:energy_flow_matrix_element_com} with the parity operator. We get
\begin{equation}
    \mathcal{P}\mathcal{E}_{\mathrm{out}}\left(n\right)\mathcal{P}^{-1} = \lim_{r\rightarrow\infty}  \int_0^\infty dt \, r^2 \left(-n^i\right) T^{0i}\left(t,r\left(- \hat{n}\right)\right)  = \mathcal{E}_{\mathrm{out}}\left(Pn\right), \label{eq:parity_EMT}
\end{equation}
where $n^\mu=\left(1,\hat{n}\right)$ and $P_{\nu}^\mu = \mathrm{diag}\left(1,-1,-1,-1\right)$. We could alternatively start from Eq.~\eqref{eq:energy-flow-fock-def} by using the action of parity on a multi-particle state. Then by computing the action of the LHS of Eq.~\eqref{eq:parity_EMT} on an m-particle parity-conjugated state
\begin{equation}
    \mathcal{P}\mathcal{E}_{\mathrm{out}}\left(n\right)\mathcal{P}^{-1}|X\rangle = \sum_{i=1}^{m} k_i^0\delta^2\left(\Omega_{Pn}-\Omega_{k_i}\right)|X\rangle ,
\end{equation}
which is consistent with the previous equation. We now conjugate the energy flow operator with time reversal. We then find
\begin{equation}
    \mathcal{T}\mathcal{E}_{\mathrm{out}}\left(n\right)\mathcal{T}^{-1}  = \lim_{r\rightarrow\infty}  \int_{-\infty}^0 dt \, r^2 \left(-n^i\right) T^{0i}\left(t,r\hat{n}\right). 
 \end{equation}
To clarify the definition of this new operator let us  also compute its action on an in-state $|X_\mathcal{T};\mathrm{in}\rangle$ composed of $m$ particles of momentum $k_{\mathcal{T},i}$ with $i=1,2,...,m$. With this we find
 \begin{equation}
\mathcal{T}\mathcal{E}_{\mathrm{out}}\left(n\right)\mathcal{T}^{-1} |X_\mathcal{T};\mathrm{in}\rangle = \sum_{i=1}^{m}k_{\mathcal{T},i}^{0}\delta^{2}\left(\Omega_{Pn}-\Omega_{k_{\mathcal{T},i}}\right)|X_{\mathcal{T}};\mathrm{in}\rangle =\mathcal{E}_{\mathrm{in}}\left(Pn\right)|X_{\mathcal{T}};\mathrm{in}\rangle \label{eq:T-reversal_conjugationEEC}
 \end{equation}
 Therefore it follows that in terms of the energy-momentum tensor the definition of the time-reversed energy flow operator is
 \begin{equation}
     \mathcal{E}_{\mathrm{in}}\left(n\right) \equiv \lim_{r\rightarrow\infty}  \int_{-\infty}^0 dt \, r^2 n^i T^{0i}\left(t,r\left(-\hat{n}\right)\right)  
 \end{equation}
 and its action on an m-particle in-state is given by
 \begin{equation}
\mathcal{E}_{\mathrm{in}}\left(n\right)|X_{\mathcal{T}};\mathrm{in}\rangle = \sum_{i=1}^{m}k_{\mathcal{T},i}^{0}\delta^{2}\left(\Omega_{n}-\Omega_{k_{\mathcal{T},i}}\right)|X_{\mathcal{T}};\mathrm{in}\rangle. \label{eq:Eflow_in}
 \end{equation}
\subsection{Proper orthochronous Lorentz transformations}
\label{app:BoostDerivation}
To derive the Lorentz transformation properties of the energy flow operator it is best to start by reformulating Eq.~\eqref{eq:energy_flow_matrix_element_com}  in a manifestly Lorentz-covariant way \cite{Belitsky_2014}:
\begin{equation}
    \mathcal{E}_{\mathrm{out}}\left( n \right) = \int_{-\infty}^{\infty} dx_{-}\lim_{x_{+}\rightarrow\infty} x_{+}^2 T^{\mu \nu}\left(x^{\mu}\right)\frac{\bar{n}_{\mu}\bar{n}_{\nu}}{n\bar{n}}, \label{eq:covariant_energy_flow}
\end{equation}
where $n^{\mu} = (1,\hat{n})$ and $\bar{n}^{\mu}=(1,-\hat{n})$ and the detector coordinate is parametrized by $x^{\mu} = x_{+}n^{\mu}+x_{-}\bar{n}^{\mu} = (t,r\hat{n})$. To understand the Lorentz transformation properties of this operator we need to consider the transformation of the detector's position. In particular we need to consider the transformation of the vector $n^{\mu}$ at $x_{+} \rightarrow \infty$ and fixed $x_{-}$. The directional unit vector is defined as:
\begin{equation}
    \hat{n}^{i} = \frac{x^{i}}{\sqrt{x_{j}x_{j}}}  
\end{equation}
We can study the transformation that it inherits from the transformation of the position vector $x'^i=\Lambda^{i}_{\ \nu}x^{\nu}$ to obtain:
\begin{equation}
   \hat{n}^{'i}= \frac{\Lambda^{i}_{\ 0}(x_{+}+x_-)+\Lambda^{i}_{\ j}(x_+-x_-)\hat{n}^{j}}{\sqrt{\Lambda^{j}_{\ 0}\Lambda^j_{0 }(x_++x_-)^2+2\Lambda^j_{0 }\Lambda^{j}_{\ i}\hat{n}^{i}(x_+^2-x_{-}^2)+\Lambda^{j}_{\ i}\Lambda^j_{\ k }\hat{n}^i \hat{n}^k(x_+-x_-)^2}}
\end{equation}
Now in the $x_+ \rightarrow \infty$ limit we get:
\begin{equation}
      \hat{n}'^{i} = \frac{\Lambda^{i}_{\ 0}+\Lambda^{i}_{\ j}\hat{n}^{j}}{\sqrt{\Lambda^{j}_{\ 0}\Lambda^j_{0 }+2\Lambda^j_{\ 0 }\Lambda^{j}_{\ i}\hat{n}^{i}+\Lambda^{j}_{\ i}\Lambda^{j}_{\ k}\hat{n}^i \hat{n}^k}} +\mathcal{O}\left(\frac{x_-}{x_+}\right)
\end{equation}
Note that the denominator can be written as the magnitude of the three-dimensional vector $\Lambda^{j}_{\ \nu}n^{\nu}$. Furthermore, since $n^{\mu}$ is light-like it follows that it squares to zero in any frame: $\Lambda^{0}_{\ \nu}n^{\nu} = |\Lambda^{j}_{\ \nu}n^{\nu}| $  ( $\Lambda^{0}_{\ \nu}n^{\nu} \geq 0 $ since $n^0=1$ and we only consider orthochronous Lorentz transformations). 
We therefore arrive at the desired transformation: 
\begin{equation}
    n^{' \mu} = \kappa\, \Lambda^{\mu}_{\ \nu}n^{\nu}+\mathcal{O}\left(\frac{x_-}{x_+}\right) \label{eq:n-sphere_boost}
\end{equation}
where $\kappa= \left(\Lambda^{0}_{\ \gamma}n^{\gamma}\right)^{-1}$ (the same factor as in \cite{Riembau:2025isw}). With this transformation the normalization of the 0th-component is not changed by the boost, i. e. $n'^\mu =\left(1,\hat{n}'\right)$. We will now drop the subleading corrections since we are working in the $x^+ \rightarrow \infty$ limit. Using now Eq.~\eqref{eq:n-sphere_boost} we can work out the transformation of $x\bar{n}$ and $xn$ by imposing that $x^{\mu}$ transforms as a contravariant Lorentz vector
\begin{equation}
    x'^{\mu} = x'_{+}n'^{\mu}+x'_{-}\bar{n}'^{\mu}= \Lambda^{\mu}_{\ \nu}x^{\nu} = \frac{1}{\kappa}x_{+}n'^\mu+\frac{1}{\bar{\kappa}}x_{-}\bar{n}'^\mu\, ,
\end{equation}
with $\bar{\kappa} = \left(\Lambda^{0}_{\ \gamma}\bar{n}^{\gamma}\right)^{-1}$. From which it follows that
\begin{equation}
    x'_{+}=\frac{x_+}{\kappa} ,\ \ \ \ \ x'_{-}=\frac{x_-}{\bar{\kappa}} \,.\label{eq:tranformation_coordinates}
\end{equation}
Substituting Eqs.~\eqref{eq:n-sphere_boost} and~\eqref{eq:tranformation_coordinates} into the covariant energy flow operator in Eq.~\eqref{eq:covariant_energy_flow} shows that the energy flow operator must transform as:
\begin{equation}
    U\left(\Lambda\right)\mathcal{E}_{\mathrm{out}}\left(n\right)   U ^\dagger \left(\Lambda\right)= \kappa^{3}\mathcal{E}_{\mathrm{out}}\left(n'\right). \label{eq:energy_flow_transform}
\end{equation}
Using this relation we can simplify the form of the different projections of Eq.~\eqref{eq:eec-tensor-def} by isolating the Lorentz-invariant pieces.  They all take the form:
\begin{equation}
    \frac{dW_{a}}{d\Omega_1d\Omega_2} = \frac{F_a(z)}{\left(n_1\cdot n_2\right)^3}, \hspace{2cm} a=0,1,2.
\end{equation}
where $z$ (defined in Eq.~\eqref{eq:covariant-z-def}) is a Lorentz-invariant combination of the available vectors. This derivation agrees with the known results in~\cite{Riembau:2025isw,Gao:2026xuq,Holguin:2026vld,Chen:2022jhb}. The power of 3 in Eq.~\eqref{eq:energy_flow_transform} is referred to as the celestial dimension of the energy flow operator.

\section{Fixed-order results in the small-acoplanarity limit}
\label{app:Derivation_zphi_fixedorder}
At tree level, we find it more convenient to compute the EEC directly in the boosted frame using its definition, rather than applying Eq.~\eqref{eq:boosted-fixed-order-integral} to the rest-frame distributions, which are proportional to $\delta(1-z)$:
\begin{align}
    \frac{d W_{a,\mathrm{tree}}}{d\Omega'_1 d\Omega'_2} = \sum_{i, j}\int d\mathrm{PS}_2 \,  E'_iE'_j \delta^{2}\left(\Omega'_{1}-\Omega'_{i} \right)\delta^{2}\left(\Omega'_{2}-\Omega'_{j} \right)|\mathcal{M}_{\mathrm{a,tree}}|^2.
\end{align}
For the real emission correction it is much more efficient to use the boost formula, which is what we did. The physical support of Eq.~\eqref{eq:boosted-fixed-order-integral} is $\max\{1-z_\theta\cosh^2\tilde{\eta},0\}<z_\phi<1$.
Away from the boundary $z_\theta=\sech^2\tilde{\eta}$, the limit
$z_\phi\to0$ can therefore only be approached on the branch
$\sech^2\tilde{\eta}<z_\theta<1$.
To motivate the origin of the relevant phase-space region expand Eq.~\eqref{eq:z-boosted-variables} for $z_\phi \rightarrow 0 $ and $\tilde{\eta}\rightarrow \eta$, which yields:
\begin{equation}
    \frac{1}{1-z} \approx \frac{z_\theta \sinh^2{\eta}}{\left(1-z_\theta\right)\left[z_\phi+\left(\tilde{\eta}-\eta\right)^2\right]}. \label{eq:boosted_hint}
\end{equation}
Therefore we expect the integrand in Eq.~\eqref{eq:boosted-fixed-order-integral} to be dominated precisely by the region  $z_\phi \sim \left(\eta-\tilde{\eta}\right)^2\ll1$. Expanding the full integrand (Eq.~\eqref{eq:boosted-fixed-order-integral}) for the $W_0$ projection (Eq.~\eqref{eq:porjectionsCOM0}) gives:
\begin{align}
    \frac{dW_0}{dz_\theta dz_\phi} &\approx -\frac{\alpha_s C_FN_c Q^4}{16\pi^3}\frac{|\csch{\eta}|\Theta\left(z_\theta-\sech^2\eta\right)}{z_\theta^{5/2}\sqrt{z_\phi\left(z_\theta \cosh^2{\eta}-1\right)}}\nonumber\\
&\quad\times\int_{-\infty}^\infty d\tilde{\eta}\, \frac{1}{{z_\phi+\left(\tilde{\eta}-\eta\right)^2}}\left[3+2\ln\left(\frac{\left(1-z_\theta\right)\left(z_\phi+\left(\tilde{\eta}-\eta\right)^2\right)}{z_\theta\sinh^2{\eta}}\right)\right].
\end{align}
The result of carrying out the integration is the second term of Eq.~\eqref{eq:endpoint_H0}. The case of $W_1$ (Eq.~\eqref{eq:porjectionsCOM1}), is simpler since the distribution in the boson center-of-mass frame has no enhancement as $z_\phi\rightarrow0$. Its leading order of magnitude is instead just fixed by the Jacobian (Eq.~\eqref{eq:angular-jacobian}) which brings a factor of $z_\phi^{-1/2}$. For $W_2$  and $\tilde{W}_{YY}$ the final result follows from observing that both these distributions are proportional to $W_0$ when $z\rightarrow1$.  The boosted result for $W_{GG}$  follows in an exactly analogous way.

Note that we performed the small-$z_\phi$ expansion at fixed $z_\theta$ and $\eta$.
The most restrictive kinematic condition arises from the approximation
\begin{equation}
\sqrt{z_\theta\sinh^2\tilde{\eta}+z_\phi-(1-z_\theta)}
\approx \sqrt{z_\theta\cosh^2\eta-1}
\end{equation}
in the Jacobian. Requiring this approximation to hold throughout the
dominant region $\tilde{\eta}-\eta=\mathcal O(\sqrt{z_\phi})$ gives
the following conditions for the validity of our expansion:
\begin{equation}
            z_{\phi}\ll 1, \hspace{1cm} z_\theta-\sech^2{\eta}\gg 2z_\theta z_\phi^{1/2}|\tanh{\eta}|\label{eq:kinematic_region}.
 \end{equation}
\section{Sum rule conventions}
\label{app:sumrules}
In this Appendix we discuss the sum rules that appear in the perturbative expansion of the factorization formula. We will compare our conventions with those of \cite{Schafer:1999kn}. See also~\cite{Meissner:2010cc}. In this appendix we suppress the hadron and parton labels, $D\equiv D_{h/q}$, $H^\perp\equiv H^\perp_{h/q}$ and $\hat{H}^{(3)}\equiv\hat{H}^{(3)}_{h/q}$. Let us start by comparing the unpolarized quark TMD fragmentation function. The definitions are related by
\begin{equation}
    D(\xi,k_\perp) = \frac{1}{\xi}D_{\mathrm{ST}}(\xi,-k_\perp/\xi).
\end{equation}
\cite{Schafer:1999kn} relates the TMDFF to the collinear fragmentation function by
\begin{equation*}
    D_{\mathrm{ST}}(\xi)=\int \frac{d^2k_\perp}{\xi^2}\, D_{\mathrm{ST}}(\xi,-k_\perp/\xi).
\end{equation*}
This is enough to obtain the sum rule in our convention
\begin{equation}
    \sum_h \int_0^1 d\xi\, \xi D(\xi) =1 \leftrightarrow     \sum_h \int_0^1 d\xi\, \xi^2D_{\mathrm{ST}}(\xi) =1 \label{eq:SumRuleUnpol}.
\end{equation}
We have an analogous relation for the Collins function.
\begin{equation}
    H^{\perp}(\xi,k_\perp) = \frac{1}{\xi}H^\perp_{\mathrm{ST}}(\xi,-k_\perp/\xi).
\end{equation}
The sum in the convention of \cite{Schafer:1999kn} is given in terms of
\begin{equation*}
    H_{\mathrm{ST}}^{\perp(1)}(\xi) = \int \frac{d^2k_\perp}{\xi^4}\,\frac{k_\perp^2}{2M_h^2}\,H_{\mathrm{ST}}^{\perp}(\xi,-k_\perp/\xi).
\end{equation*}
To relate this to our convention first note that by using both Eq.~\eqref{eq:CollinsOPE} and Eq.~\eqref{eq:CollinsFT} to lowest order in  $b\Lambda_{\mathrm{QCD}}$ we find that the twist-3 object in the first term of the OPE in Eq.~\eqref{eq:CollinsOPE} is related to the Collins function by
\begin{equation}
\hat{H}^{(3)}(\xi) =\frac{-1}{\xi^2M_h}\int d^2k_\perp k_\perp^2 H^\perp(\xi,k_\perp),\label{eq:Twist3-CollinsOPEpiece}
\end{equation}
from which it follows that $\hat{H}^{(3)}(\xi) =-2M_h\xi H^{\perp(1)}_{\mathrm{ST}}(\xi)$ and therefore the corresponding sum rule conventions are related by
\begin{equation}
    \sum_h  \int_0^1 d\xi \,\xi\hat{H}^{(3)}(\xi) = 0 \leftrightarrow \sum_h M_h\int_0^1 d\xi\, \xi^2 H^{\perp(1)}_{\mathrm{ST}}(\xi) = 0. \label{eq:CollinsSumRule}
\end{equation}
\section{Verifying perturbative limit of factorization formula to one loop}
\label{app:pert_check_fact}
\subsection{Quark-initiated EEC }
In this appendix we match the factorization formula in Eq.~\eqref{eq:ZEEC_polarization_decomp} to the fixed-order results in Eqs.~\eqref{eq:endpoint_H0}--\eqref{eq:endpoint_H2}. We will drop the terms depending on the Collins function since they vanish in perturbation theory. Using $e_x^\mu e_x^\nu+e_y^\mu e_y^\nu+e_z^\mu e_z^\nu=-g_{\perp}^{\mu\nu}$ it follows that 
\begin{equation}
    g_{\perp,\mu\nu}\left(e_x^\mu e_x^\nu+e_y^\mu e_y^\nu\right)= -2,
\end{equation}\begin{equation}
    C_{\mu\nu}\left(e_x^\mu e_x^\nu+e_y^\mu e_y^\nu\right)= 1-\left(e_c \cdot e_z\right)^2 \approx 1 ,
\end{equation}\begin{equation}
    D_{\mu\nu}\left(e_x^\mu e_x^\nu+e_y^\mu e_y^\nu\right)= 1-\left(e_d \cdot e_z\right)^2 \approx 0.
\end{equation}
The approximations hold when the detectors are back-to-back. Small $z_\phi$ alone enforces opposite azimuths, but does not require the
pair to be back-to-back in the boson rest frame. The leading fixed-order
enhancement, however, is generated by the joint region
$(\tilde\eta-\eta)^2\sim z_\phi\ll1$, in which $1-z=\mathcal O(z_\phi)$.
We may therefore evaluate the tensor contractions in the back-to-back
limit for the leading-power matching in $z_\phi$. At this point we can already verify that $W_1=0$ and $W_0=-2W_2$ which match the fixed-order results.  Therefore, dropping the couplings in Eq.~\eqref{eq:ZEEC_polarization_decomp} we get
\begin{align}
     \frac{d W_{0}}{d\zph\,d\zth}=& \frac{N_cQ^4}{8\pi} H(\mu,Q)\mathcal{K}\left(z_\theta;\eta\right)
    \nonumber \\
    &\times\int d{l}_{\phi} \ \delta\left(z_{\phi}-\frac{l_\phi ^2}{4p_{T}^2}\right) \int\frac{db}{2\pi} e^{-ibl_\phi } J_{q}\left(b, \mu, \zeta_1/\nu^2\right) J_{\bar{q}}\left( b, \mu, \zeta_2/\nu^2\right) S_{n_1 n_2}\left(b, \mu, \nu\right)\,.
    \label{eq:factorization-qy-by}
\end{align}
Here we will only keep the perturbative expansion of the jet functions in Eq.~\eqref{eq:UnpolQuarkJetfunction} the same way as it was done for Section~\ref{sec:InclusiveEEC}. We grab the perturbative results for the jet, soft and hard functions from \cite{Moult:2018jzp} to order $\mathcal{O}(\alpha_{s})$. Even though the factorization presented in this paper is one-dimensional, the dependence of the jet and soft functions on $b$ is isotropic and therefore we can reuse the results computed for a two-dimensional factorization as in \cite{Moult:2018jzp}.
\begin{align}
H(\mu,Q)
&=
1+\frac{\alpha_s}{4\pi} C_F
\left[
-2\ln^2\!\left(\frac{\mu^2}{Q^2}\right)
-6\ln\!\left(\frac{\mu^2}{Q^2}\right)
+14\zeta(2)-16
\right]
+\mathcal O(\alpha_s^2),
\label{eq:hard-one-loop}
\\[1em]
J_1\!\left(b,\mu,\frac{\zeta_1}{\nu^2}\right)
&=
1+\frac{\alpha_s}{4\pi}  C_F
\left[
\ln\!\left(\frac{b^2\mu^2}{b_0^2}\right)
\left(
3-2\ln\!\frac{\zeta_1}{\nu^2}
\right)
+4-8\zeta(2)
\right]
+\mathcal O(\alpha_s^2),
\label{eq:jet1-one-loop}
\\[1em]
J_2\!\left(b,\mu,\frac{\zeta_2}{\nu^2}\right)
&=
1+\frac{\alpha_s}{4\pi}  C_F
\left[
\ln\!\left(\frac{b^2\mu^2}{b_0^2}\right)
\left(
3-2\ln\!\frac{\zeta_2}{\nu^2}
\right)
+4-8\zeta(2)
\right]
+\mathcal O(\alpha_s^2),
\label{eq:jet2-one-loop}
\\[1em]
S_{n_1n_2}(b,\mu,\nu)
&=
1+\frac{\alpha_s}{4\pi}  C_F
\left[
2\ln^2\!\left(\frac{b^2\mu^2}{b_0^2}\right)
-4\ln\!\left(\frac{b^2\mu^2}{b_0^2}\right)
\ln\!\left(
\frac{b^2\nu^2}{b_0^2}z_\theta
\right)
-2\zeta(2)
\right]
+\mathcal O(\alpha_s^2).
\label{eq:soft-one-loop}
\end{align}
The product can be packaged in the following way
\begin{align}
H\,J_1\,J_2\,S_{n_1n_2}
&=
1+\frac{\alpha_s(\mu)C_F}{4\pi}
\Bigg\{
-2\ln^2\!\left(\frac{b^2Q^2}{b_0^2}\right)
\nonumber\\
&\quad
+
\left[
6
-4\ln\!\left(
\frac{\sqrt{\zeta_1\zeta_2}}{Q^2}
z_\theta
\right)
\right]
\ln\!\left(\frac{b^2Q^2}{b_0^2}\right)
\nonumber\\
&\quad
-4
\ln\!\left(
\frac{\sqrt{\zeta_1\zeta_2}}{Q^2}
z_\theta
\right)
\ln\!\left(\frac{\mu^2}{Q^2}\right)
-8-4\zeta(2)
\Bigg\}
+\mathcal O(\alpha_s^2).
\label{eq:product-arbitrary-zeta}
\end{align}
Since $\zeta_i^{1/2}=2E_{i}$, then $\frac{\sqrt{\zeta_1\zeta_2}z_{\theta}}{Q^2}=1$, such that the product reduces to:
\begin{align}
H\,J_1\,J_2\,S_{n_1n_2}
&=
1+\frac{\alpha_s(\mu)C_F}{4\pi}
\Bigg\{
-2\ln^2\!\left(\frac{b^2Q^2}{b_0^2}\right)
+
6
\ln\!\left(\frac{b^2Q^2}{b_0^2}\right)
-8-4\zeta(2)
\Bigg\}
+\mathcal O(\alpha_s^2).
\label{eq:product-natural-zeta}
\end{align}
We will have to compute the integral transformations:
\begin{equation}
    K\left[F(b)\right]=\int d{l}_{\phi} \ \delta\left(z_{\phi}-\frac{l_{\phi}^2}{4p_{T}^2}\right) \int\frac{db}{2\pi} e^{-ibl_\phi }F(b).
    \label{eq:K-transform-def}
\end{equation}
The relevant terms give:
\begin{align}
&K\left[1\right]=\delta\left(z_{\phi}\right),
\label{eq:K-one}
\\[1em]
&K\left[\ln{\left(\frac{b^2Q^2}{b_0^2}\right)}\right]=
\delta\left(z_{\phi}\right)\ln{\left(\frac{Q^2}{16p_T^2}\right)}-\left[\frac{1}{z_{\phi}}\right]_{+},
\label{eq:K-log}
\\[1em]
&K\left[\ln^2{\left(\frac{b^2Q^2}{b_0^2}\right)}\right]=\delta\left(z_{\phi}\right)\left(\ln{^2\left(\frac{Q^2}{16p_T^2}\right)}+\frac{\pi^2}{3}\right)-2\left[\frac{1}{z_{\phi}}\right]_{+}\ln{\left(\frac{Q^2}{16p_T^2}\right)}+2\left[\frac{\ln{z_{\phi}}}{z_{\phi}}\right]_{+}.
\label{eq:K-log-squared}
\end{align}

Finally substituting Eq.~\eqref{eq:pT-ztheta-eta} we find
\begin{equation}
\begin{aligned}
 \frac{dW_{0}}{d\zph\,d\zth}=& \frac{N_cQ^4}{8\pi}\mathcal{K}(z_\theta;\eta)\Bigg\{ \delta\left(z_{\phi}\right)
 \\[1em]
&\quad+\frac{\alpha_s(\mu) C_F}{2\pi}
\Bigg[\delta\left(z_{\phi}\right)\left(
-4-4\zeta(2)
-3\ln\left(\frac{4(1-z_{\theta})}{z_{\theta}\sinh^2\eta}\right)
-\ln^2\left(\frac{4(1-z_{\theta})}{z_{\theta}\sinh^2\eta}\right)
\right)
\\[1em]
&\qquad
-\left[\frac{1}{z_{\phi}}\right]_{+}\left(
3+2\ln\left(\frac{4(1-z_{\theta})}{z_{\theta}\sinh^2\eta}\right)
\right)
-2\left[\frac{\ln z_{\phi}}{z_{\phi}}\right]_{+}\Bigg]\Bigg\}.
\end{aligned}
\label{eq:endpoint-before-pT-substitution}
\end{equation}
Comparing to Eq.~\eqref{eq:endpoint_H0} we see that this expansion agrees with the leading Born term and the leading real emission. Eq.~\eqref{eq:endpoint-before-pT-substitution} regularizes the endpoints with a plus distribution and also gives the one-loop endpoint term, which we did not see in the fixed-order expressions since we did not include the virtual corrections there.

\subsection{Gluon-initiated EEC}
In this subsection we will match the factorization formula in Eq.~\eqref{eq:Hggfactfinal} to the fixed-order result in Eq.~\eqref{eq:HiggsGluons}.
In this case it is important to note that the contribution from the linearly polarized gluon EEC jet function is non-vanishing in perturbation theory, however its perturbative expansion only begins at $\mathcal{O}(\alpha_s)$~\cite{Ebert:2020sfi} and therefore we discard it in this expansion. We begin by matching the unpolarized gluon TMDFF to its collinear fragmentation function:
\begin{align}
D_{h/g}(\xi,b,\mu,\zeta)&=\frac{1}{\xi}\sum_{j}\int_\xi^1\frac{d\tau}{\tau}\;C_{j/g}\Big(\frac{b}{\tau},\tau,\mu,\zeta\Big)\,D_{h/j}\Big(\frac{\xi}{\tau},\mu\Big),
\end{align}
from which it follows that
\begin{equation}
J_g\left(b,\mu,\zeta/\nu^2\right)\equiv\sum_{j}\int_0^1d\tau\,\tau \,C_{j/g}\left(\frac{b}{\tau},\tau,\mu,\zeta\right)\sum_h\int _0^1d\xi \, \xi D_{h/j}\Big({\xi},\mu\Big),
\end{equation}
where the sum rule gets rid of the collinear distributions. Then in perturbation theory the gluon EEC jet function is given by \cite{Luo:2019bmw}
\begin{align}
J_g\big(b,\mu,\zeta/\nu^2\big)=1&+\frac{\alpha_s(\mu)}{4\pi}\Big[-2C_A\,\ln\!\left(\frac{b^2\mu^2}{b_0^2}\right)\ln\!\left(\frac{\zeta}{\nu^2}\right)+\beta_0\,\ln\!\left(\frac{b^2\mu^2}{b_0^2}\right)\nonumber\\
&\qquad\quad+C_A\left(\frac{65}{18}-8\zeta_2\right)-\frac{5}{9}\,T_F N_f\Big]+\mathcal{O}(\alpha_s^2)\,.
\end{align}
We take the one-loop Higgs-to-two-gluon hard function from~\cite{Ahrens:2009cxz}. For the soft function we use the adjoint soft function $S^A_{n_1n_2}$ appearing in the factorization of the Higgs EEC in the back-to-back limit~\cite{Ebert:2020sfi}, which in the exponential regulator is given by the TMD soft function of~\cite{Li:2016axz} with $C_a=C_A$. As in~\cite{Gao:2019ojf}, the rapidity scale is shifted, $\nu^2 \rightarrow\nu^2 z_\theta$:
\begin{align}
H^{G}(\mu,Q)&=1+\frac{\alpha_s(\mu)}{4\pi}\,C_A\left[-2\ln^2\!\left(\frac{\mu^2}{Q^2}\right)+14\zeta(2)\right]+\mathcal{O}(\alpha_s^2)\,,\\
S^A_{n_1n_2}(b,\mu,\nu)&=1+\frac{\alpha_s(\mu)}{4\pi}\,C_A\left[2\ln^2\!\left(\frac{b^2\mu^2}{b_0^2}\right)-4\ln\!\left(\frac{b^2\mu^2}{b_0^2}\right)\ln\!\left(\frac{b^2\nu^2}{b_0^2}\,z_\theta\right)-2\zeta(2)\right]+\mathcal{O}(\alpha_s^2)\,.
\end{align}
We are now ready to expand the factorization formula to first non-trivial order. Integrating Eq.~\eqref{eq:factorization_Hgg} over $\bar\phi$ and stripping the effective coupling $C_g^2$, the factorization formula for the hadronic function of the current $G^b_{\mu\nu}G^{b\,\mu\nu}$ as in Eq.~\eqref{eq:HiggsGluonFO} reads
\begin{align}
\frac{d\tilde{W}_{GG}}{dz_\phi\,dz_\theta}={}&\left(N_c^2-1\right)\frac{Q^6}{8\pi}\,H^{G}(\mu,Q)\,\mathcal K(z_\theta;\eta)\int dl_\phi\,\delta\!\left(z_\phi-\frac{l_\phi^2}{4p_T^2}\right)\nonumber\\
&\times\int\frac{db}{2\pi}\,e^{-ibl_\phi}\,S^A_{n_1n_2}(b,\mu,\nu)\,J_g\!\left(b,\mu,\zeta_1/\nu^2\right)J_g\!\left(b,\mu,\zeta_2/\nu^2\right)\,,
\end{align}
Following the steps of the previous section we arrive at
\begin{align}
\frac{d\tilde{W}_{GG}}{dz_\phi\,dz_\theta}={}&\left(N_c^2-1\right)\frac{Q^6}{8\pi}\,\mathcal K(z_\theta;\eta)\Bigg\{\delta(z_\phi)+\frac{\alpha_s(\mu)}{2\pi}\Bigg[\delta(z_\phi)\Bigg(C_A\left(\frac{65}{18}-4\zeta(2)\right)-\frac{5}{9}T_FN_f\nonumber\\
&\qquad-\beta_0\ln\frac{4(1-z_\theta)}{z_\theta\sinh^2\eta}-C_A\ln^2\frac{4(1-z_\theta)}{z_\theta\sinh^2\eta}+\beta_0\ln\frac{\mu^2}{Q^2}\Bigg)\nonumber\\
&\qquad-\left[\frac{1}{z_\phi}\right]_+\left(\beta_0+2C_A\ln\frac{4(1-z_\theta)}{z_\theta\sinh^2\eta}\right)-2C_A\left[\frac{\ln z_\phi}{z_\phi}\right]_+\Bigg]\Bigg\}+\mathcal O(\alpha_s^2)\,. \label{eq:gluon-endpoint-expansion}
\end{align}
Similar to the quark case, we see that this expansion agrees with the leading Born term and the leading real emission in Eq.~\eqref{eq:HiggsGluonFO}, where Eq.~\eqref{eq:gluon-endpoint-expansion} regularizes the endpoints with a plus distribution and also gives the one-loop endpoint term.

Finally, the full $\alpha_s$ dependence of the $H\to gg$ rate is obtained by multiplying by $C_g^2(\mu)$, where in the heavy-top limit~\cite{Ahrens:2009cxz}
\begin{align}
C_g(\mu)=-\frac{\alpha_s(\mu)}{12\pi v}\left[1+\frac{\alpha_s(\mu)}{4\pi}\left(5C_A-3C_F\right)+\mathcal O(\alpha_s^2)\right].
\end{align}
Since $\alpha_s^2(\mu)=\alpha_s^2(Q)\left[1-\frac{\alpha_s}{2\pi}\beta_0\ln\frac{\mu^2}{Q^2}+\mathcal O(\alpha_s^2)\right]$, the term $\beta_0\ln(\mu^2/Q^2)$ in the coefficient of $\delta(z_\phi)$ cancels.

\section{Density matrices for photon-tagged Z hadronic decay in $e^+e^- \rightarrow  Z\gamma$ }

 \subsection{Photon-tagged Z density matrix }
  \label{app:ZDensityMatrix}
 To obtain the $Z$ boson density matrix necessary in Eq.~\eqref{eq:Z_fact} we take the helicity amplitudes given in~\cite{Ellis:2019zex} and sum over the electron and photon helicities (we also normalize it such that it has unit trace)
 \begin{equation}
\rho_{\lambda\lambda'}(\theta_\gamma)=\frac{\sum_{s,\lambda_{\gamma}}\mathcal T^{s,-s}_{\lambda,\lambda_\gamma}\left(\theta_\gamma\right)\,\left(\mathcal T^{s,-s}_{\lambda'\lambda_\gamma}\left(\theta_\gamma\right)\right)^*}{\sum_{s,\lambda_\gamma,\lambda}\big|\mathcal T^{s,-s}_{\lambda\lambda_\gamma}\left(\theta_\gamma\right)\big|^2},\qquad
s=\pm\frac{1}{2},\lambda_\gamma = \pm. \label{eq:ZdensitymatrixExplicit}
\end{equation}
In the polarization convention used here, the production density matrix is real and symmetric. Together with unit trace, this leaves five independent real components. Defining the common normalization factor
\begin{equation}
\mathcal{N}
\equiv
\frac{1}{
4(c_L^2+c_R^2)
\bigl[\cosh^2\eta+\sinh^2\eta\cos^2\theta_\gamma\bigr]},
\end{equation}
the entries are
\begin{equation}
\begin{aligned}
\rho_{++}(\theta_\gamma)
&=\mathcal{N}\cosh(2\eta)
  \bigl[c_L^2(1+\cos\theta_\gamma)^2
       +c_R^2(1-\cos\theta_\gamma)^2\bigr],
\\
\rho_{--}(\theta_\gamma)
&=\mathcal{N}\cosh(2\eta)
  \bigl[c_L^2(1-\cos\theta_\gamma)^2
       +c_R^2(1+\cos\theta_\gamma)^2\bigr],
\\
\rho_{+0}(\theta_\gamma)
&=\sqrt{2}\mathcal{N}\cosh\eta\,\sin\theta_\gamma
  \bigl[c_L^2(1+\cos\theta_\gamma)
       -c_R^2(1-\cos\theta_\gamma)\bigr],
\\
\rho_{+-}(\theta_\gamma)
&=\mathcal{N}(c_L^2+c_R^2)\sin^2\theta_\gamma,
\\
\rho_{0-}(\theta_\gamma)
&=\sqrt{2}\mathcal{N}\cosh\eta\,\sin\theta_\gamma
  \bigl[c_L^2(1-\cos\theta_\gamma)
       -c_R^2(1+\cos\theta_\gamma)\bigr].
\end{aligned}
\label{eq:Zdensitymatrix}
\end{equation}
Here \(c_L=-\tfrac12+\sin^2\theta_W\) and
\(c_R=\sin^2\theta_W\) are the left- and right-handed
\(Zee\) couplings with the common electroweak factor removed and $\eta = \ln\left(\sqrt{s}/M_Z \right)$ is the rapidity of the $Z$ boson which is fixed by the center-of-mass energy of the electron-positron collision. In the helicity  amplitudes of Eq.~\eqref{eq:Zdensitymatrix} the $Z$ boson's helicity is quantized in the helicity frame, i.e. with respect to the $Z$ boson's momentum in the $e^+e^-$ center-of-mass frame, while the polarization vectors in Eq.~\eqref{eq:Z_fact} are quantized with respect to the quark momentum in the boosted frame. Therefore we need to construct the Wigner spin-1 change of basis matrix given by
\begin{equation}
    D^{(1)}_{\lambda,a} = -\left( \tilde{\epsilon}_\lambda \right)^* \cdot \epsilon_a, \hspace{2cm} a,\lambda = +,-,0.
\end{equation}
where $\tilde{\epsilon}_\lambda$ are the polarization vectors quantized in the $Z$ boson helicity frame and $\epsilon_a$ are the polarization vectors defined from Eqs.~\eqref{eq:polarization_z}--\eqref{eq:polarization_y}. This is most easily obtained in the $Z$ boson COM frame with the coordinate
\(\hat z\)-axis along the original \(Z\) flight direction in the
collider center-of-mass frame. Let $\varphi$  be the azimuthal angle, measured from the chosen $\hat{x}$ axis about the $Z$ boson flight direction, to the direction opposite the incoming electron’s transverse momentum
\begin{equation}
    \hat{p}_{e^-,\perp} = - \cos{\varphi}\,\hat{x} - \sin{\varphi}\,\hat{y}\,. \label{eq:varphi_def}
\end{equation}The production
polarization vectors in these coordinates are
\begin{equation}
\begin{aligned}
\widetilde{\epsilon}_{\pm}^{\mu}
  &= \frac{e^{\mp i\varphi}}{\sqrt{2}}(0,\mp 1,-i,0),\\
\widetilde{\epsilon}_{0}^{\mu}
  &= (0,0,0,1).
\end{aligned}
\end{equation}
Let \(\theta^*\) (Eq.~\eqref{eq:sinCOM}) and \(\phi_1\) (Eq.~\eqref{eq:boosted-born-momenta}) denote the quark polar and azimuthal angles
in the same coordinates. The quark-axis polarization vectors are
\begin{equation}
\begin{aligned}
\epsilon_{+}^{\mu}
  &= \frac{1}{\sqrt{2}}\bigl(
     0,\,
     -\cos\theta^*\cos\phi_1+i\sin\phi_1,\,
     -\cos\theta^*\sin\phi_1-i\cos\phi_1,\,
     \sin\theta^*
     \bigr),\\
\epsilon_{0}^{\mu}
  &= \bigl(
     0,\,
     \sin\theta^*\cos\phi_1,\,
     \sin\theta^*\sin\phi_1,\,
     \cos\theta^*
     \bigr),\\
\epsilon_{-}^{\mu}
  &= -\frac{1}{\sqrt{2}}\bigl(
     0,\,
     -\cos\theta^*\cos\phi_1-i\sin\phi_1,\,
     -\cos\theta^*\sin\phi_1+i\cos\phi_1,\,
     \sin\theta^*
     \bigr).
\end{aligned}
\end{equation}
Then the Wigner spin-1 matrix is given by
\begin{equation}
D^{(1)}_{\lambda a}
= -(\widetilde{\epsilon}_{\lambda})^{*}\cdot\epsilon_a
= \frac{1}{2}
\setlength{\arraycolsep}{4pt}
\begin{pmatrix}
(1+\cos\theta^{*})e^{i(\varphi-\phi_1)}
&
-\sqrt{2}\sin\theta^{*}\,e^{i(\varphi-\phi_1)}
&
(1-\cos\theta^{*})e^{i(\varphi-\phi_1)}
\\[4pt]
\sqrt{2}\sin\theta^{*}
&
2\cos\theta^{*}
&
-\sqrt{2}\sin\theta^{*}
\\[4pt]
(1-\cos\theta^{*})e^{-i(\varphi-\phi_1)}
&
\sqrt{2}\sin\theta^{*}\,e^{-i(\varphi-\phi_1)}
&
(1+\cos\theta^{*})e^{-i(\varphi-\phi_1)}
\end{pmatrix}.
\end{equation}
Conjugating Eq.~\eqref{eq:ZdensitymatrixExplicit} with this, we obtain the $Z$ boson density matrix $\rho'$ which enters the factorization formula in Eq.~\eqref{eq:Z_fact} . We display two diagonal entries and the three upper-triangular entries. The remaining entries follow from unit trace and hermiticity. Note that in writing these entries we use the Born-level approximation $\phi_1 \approx \ \bar{\phi}-\pi/2$ as discussed around Eq.~\eqref{eq:azimuth_Modulation}.
\begin{align}
\rho'_{++}
&= \frac{(1+\cos\theta^*)^2}{4}\,\rho_{++}(\theta_\gamma)
 + \frac{(1-\cos\theta^*)^2}{4}\,\rho_{--}(\theta_\gamma)
 \notag\\
&\quad
 + \frac{\sin^2\theta^*}{2}
   \Bigl[\rho_{00}(\theta_\gamma)
   - \rho_{+-}(\theta_\gamma)
     \cos\!\bigl[2(\bar{\phi}-\varphi)\bigr]\Bigr]
 \notag\\
&\quad
 + \frac{\sin\theta^*}{\sqrt{2}}
   \Bigl[
     (1+\cos\theta^*)\,\rho_{+0}(\theta_\gamma)
     +(1-\cos\theta^*)\,\rho_{0-}(\theta_\gamma)
   \Bigr]\sin(\bar{\phi}-\varphi),
\\[6pt]
\rho'_{--}
&= \frac{(1-\cos\theta^*)^2}{4}\,\rho_{++}(\theta_\gamma)
 + \frac{(1+\cos\theta^*)^2}{4}\,\rho_{--}(\theta_\gamma)
 \notag\\
&\quad
 + \frac{\sin^2\theta^*}{2}
   \Bigl[\rho_{00}(\theta_\gamma)
   - \rho_{+-}(\theta_\gamma)
     \cos\!\bigl[2(\bar{\phi}-\varphi)\bigr]\Bigr]
 \notag\\
&\quad
 - \frac{\sin\theta^*}{\sqrt{2}}
   \Bigl[
     (1-\cos\theta^*)\,\rho_{+0}(\theta_\gamma)
     +(1+\cos\theta^*)\,\rho_{0-}(\theta_\gamma)
   \Bigr]\sin(\bar{\phi}-\varphi),
\displaybreak[0]\\[6pt]
\rho'_{+0}
&= \frac{\sin\theta^*}{2\sqrt{2}}
   \Bigl[
     \rho_{--}(\theta_\gamma)-\rho_{++}(\theta_\gamma)
     -\cos\theta^*\bigl[1-3\rho_{00}(\theta_\gamma)\bigr]
   \Bigr]
 \notag\\
&\quad
 - \frac{\sin\theta^*}{\sqrt{2}}\,\rho_{+-}(\theta_\gamma)
   \Bigl[
     \cos\theta^*\cos\!\bigl[2(\bar{\phi}-\varphi)\bigr]
     +i\sin\!\bigl[2(\bar{\phi}-\varphi)\bigr]
   \Bigr]
 \notag\\
&\quad
 + \frac{\rho_{+0}(\theta_\gamma)+\rho_{0-}(\theta_\gamma)}{2}
   \Bigl[
     \cos\theta^*\sin(\bar{\phi}-\varphi)
     -i\cos(\bar{\phi}-\varphi)
   \Bigr]
 \notag\\
&\quad
 + \frac{\rho_{+0}(\theta_\gamma)-\rho_{0-}(\theta_\gamma)}{2}
   \Bigl[
     \cos(2\theta^*)\sin(\bar{\phi}-\varphi)
     -i\cos\theta^*\cos(\bar{\phi}-\varphi)
   \Bigr],
\displaybreak[0]\\[6pt]
\rho'_{+-}
&= \frac{\sin^2\theta^*}{4}
   \bigl[1-3\rho_{00}(\theta_\gamma)\bigr]
 \notag\\
&\quad
 - \rho_{+-}(\theta_\gamma)
   \Bigl[
     \frac{1+\cos^2\theta^*}{2}
       \cos\!\bigl[2(\bar{\phi}-\varphi)\bigr]
     +i\cos\theta^*
       \sin\!\bigl[2(\bar{\phi}-\varphi)\bigr]
   \Bigr]
 \notag\\
&\quad
 + \frac{\sin\theta^*}{\sqrt{2}}
   \bigl[\rho_{0-}(\theta_\gamma)-\rho_{+0}(\theta_\gamma)\bigr]
   \Bigl[
     \cos\theta^*\sin(\bar{\phi}-\varphi)
     -i\cos(\bar{\phi}-\varphi)
   \Bigr],
\displaybreak[0]\\[6pt]
\rho'_{0-}
&= \frac{\sin\theta^*}{2\sqrt{2}}
   \Bigl[
     \rho_{--}(\theta_\gamma)-\rho_{++}(\theta_\gamma)
     +\cos\theta^*\bigl[1-3\rho_{00}(\theta_\gamma)\bigr]
   \Bigr]
 \notag\\
&\quad
 + \frac{\sin\theta^*}{\sqrt{2}}\,\rho_{+-}(\theta_\gamma)
   \Bigl[
     \cos\theta^*\cos\!\bigl[2(\bar{\phi}-\varphi)\bigr]
     +i\sin\!\bigl[2(\bar{\phi}-\varphi)\bigr]
   \Bigr]
 \notag\\
&\quad
 + \frac{\rho_{+0}(\theta_\gamma)+\rho_{0-}(\theta_\gamma)}{2}
   \Bigl[
     \cos\theta^*\sin(\bar{\phi}-\varphi)
     -i\cos(\bar{\phi}-\varphi)
   \Bigr]
 \notag\\
&\quad
 - \frac{\rho_{+0}(\theta_\gamma)-\rho_{0-}(\theta_\gamma)}{2}
   \Bigl[
     \cos(2\theta^*)\sin(\bar{\phi}-\varphi)
     -i\cos\theta^*\cos(\bar{\phi}-\varphi)
   \Bigr].
\end{align}

The  factorization formula in Eq.~\eqref{eq:Z_fact} is written in terms of the sum of the diagonal transverse-helicity terms $\rho_{++}'+\rho_{--}'$ and the interference terms $\rho_{+-}'+\rho_{-+}'$. Furthermore, the factorization formula only keeps terms in this expression proportional to 1, $\sin^2{\theta^*}$ and $\cos{^2\theta^*}$. To see this observe that $\cos{\theta^*}=\tanh{(\left(\eta_1-\eta_2\right)/2)}$ where $\eta_1$ and $\eta_2$ are the rapidities of the detected hard pair at parton level and 
\begin{equation}
z_\theta = \left(1+\sinh^2\eta \sech^2{((\eta_1-\eta_2)/2)}\right)^{-1} \,.    
\end{equation} 
Therefore $z_\theta$ determines $|\eta_1-\eta_2|$ but not its sign, and only terms even under the exchange of the rapidities survive in the factorization formula. The terms proportional to $\sin\theta^*$ appear in $\rho_{++}'+\rho_{--}'$ and $\rho_{+-}'+\rho_{-+}'$ only multiplied by $\cos\theta^*$. They are therefore odd under the exchange of the rapidities and are averaged out.

 \subsection{Quark-antiquark joint density matrix }
  \label{app:qqbarDensityMatrix}
At this point it is straightforward to compute the density matrix of the quark-antiquark pair that comes from the Z decay by using
\begin{equation}
    \rho_{q\bar{q}} = \frac{\mathcal{M}_{Z\rightarrow q\bar{q}} \, \rho' \, \mathcal{M}^\dagger_{Z\rightarrow q\bar{q}}}{\mathrm{Tr}\left[\mathcal{M}_{Z\rightarrow q\bar{q}} \, \rho' \, \mathcal{M}^\dagger_{Z\rightarrow q\bar{q}}\right]}, \label{eq:qqdensitymatrix}
\end{equation}where $M_{Z\rightarrow q\bar{q}}$ is a matrix in the helicity of the $q\bar{q}$ pair and the helicity of the parent $Z$ boson and $\rho'$ is the $Z$ boson density matrix computed in the previous subsection. The helicities of the pair $h\bar{h} \in \{ ++,+-,-+,--\}$ are defined in the quark helicity frame specified by the polarization constructed from Eqs.~\eqref{eq:polarization_z}--\eqref{eq:polarization_y}. In the massless quark limit the Z current produces opposite helicity pairs only. Therefore with the $Z$ spin quantized along the quark momentum, helicity conservation implies that the only nonzero entries of the amplitudes in Eq.~\eqref{eq:qqdensitymatrix} are $M_{+-,+} \propto g^q_R$ and $M_{-+,-} \propto g^q_L$. The resulting non-zero block (in the $\{+-,-+ \}$ basis) of the quark-antiquark density matrix is given by
\begin{equation}
\rho_{q\bar{q}}
=
\frac{1}{(g_R^q)^2\rho'_{++}+(g_L^q)^2\rho'_{--}}
\begin{pmatrix}
(g_R^q)^2\rho'_{++} & g_R^q g_L^q\,\rho'_{+-} \\
g_R^q g_L^q\,\rho'_{-+} & (g_L^q)^2\rho'_{--}
\end{pmatrix}.
\end{equation}
To run the simulations of the Collins effect for Fig.~\ref{fig:RatioPlots} we used the package \textsc{StringSpinner}~\cite{Kerbizi:2026iws,Kerbizi:2021pzn,Kerbizi:2023luv,Kerbizi:2024vpd} which takes as input the spin-correlation matrix $C_{ij} = \mathrm{Tr}\left[\rho_{q\bar{q}}\left(\sigma_i\otimes \sigma_j\right)\right]$. The non-zero entries are given by
\begin{align}
C_{00} &= 1, \qquad C_{zz} = -1, \\
C_{z0} &= -C_{0z}
= \frac{(g_R^q)^2\rho'_{++} - (g_L^q)^2\rho'_{--}}
       {(g_R^q)^2\rho'_{++} + (g_L^q)^2\rho'_{--}}, \\
C_{xx} &= C_{yy}
= \frac{g_R^q g_L^q\bigl(\rho'_{+-}+\rho'_{-+}\bigr)}
       {(g_R^q)^2\rho'_{++} + (g_L^q)^2\rho'_{--}}, \\
C_{xy} &= -C_{yx}
= \frac{-i g_R^q g_L^q\bigl(\rho'_{+-}-\rho'_{-+}\bigr)}
       {(g_R^q)^2\rho'_{++} + (g_L^q)^2\rho'_{--}}.
\end{align}
\bibliographystyle{JHEP-2modlong.bst}
\bibliography{references}

\end{document}